\documentclass[aps,Preprint,superscriptaddress]{revtex4-2}
\usepackage[utf8]{inputenc}
\usepackage{physics}
\usepackage{mathrsfs}
\usepackage[colorlinks=true,citecolor=blue,linkcolor=blue]{hyperref}
\usepackage[english]{babel}
\usepackage[T1]{fontenc}
\usepackage[left=2cm,right=2cm,top=2cm,bottom=2cm]{geometry}
\usepackage{amsmath,amssymb,amsthm,dsfont}
\usepackage{amsfonts}
\usepackage{graphicx}
\usepackage{subfig}
\usepackage{wrapfig}
\usepackage{slashed}
\usepackage{fancyhdr}
\selectfont
\usepackage{babel}

\begin{document}


\title{Elastic scattering of a muon neutrino by an electron in the presence of a circularly polarized laser field}


\author{S. El Asri}
\affiliation{Sultan Moulay Slimane University, Polydisciplinary Faculty, Laboratory of Research in Physics $\&$ Engineering Sciences, Team of Modern and Applied Physics, Beni Mellal, 23000, Morocco.}
\author{S. Mouslih}
\affiliation{ Faculty of Sciences and Techniques,
		Laboratory of Materials Physics (LMP),
		Beni Mellal, 23000, Morocco.}
		\affiliation{Sultan Moulay Slimane University, Polydisciplinary Faculty, Laboratory of Research in Physics $\&$ Engineering Sciences, Team of Modern and Applied Physics, Beni Mellal, 23000, Morocco.}
\author{M. Jakha}
\affiliation{Sultan Moulay Slimane University, Polydisciplinary Faculty, Laboratory of Research in Physics $\&$ Engineering Sciences, Team of Modern and Applied Physics, Beni Mellal, 23000, Morocco.}
\author{B. Manaut}
\affiliation{Sultan Moulay Slimane University, Polydisciplinary Faculty, Laboratory of Research in Physics $\&$ Engineering Sciences, Team of Modern and Applied Physics, Beni Mellal, 23000, Morocco.}
\author{Y. Attaourti}
\affiliation{High Energy Physics and Astrophysics Laboratory, FSSM, Cadi Ayyad University, Marrakesh, Morocco.}
\author{S. Taj}
\email[]{s.taj@usms.ma}
\affiliation{Sultan Moulay Slimane University, Polydisciplinary Faculty, Laboratory of Research in Physics $\&$ Engineering Sciences, Team of Modern and Applied Physics, Beni Mellal, 23000, Morocco.}
\author{R. Benbrik}
\affiliation{Polydisciplinary Faculty, Laboratoire de Physique Fondamentale et Appliquée, Sidi Bouzid, B.P. 4162 Safi, Morocco.}


\date{\today}

\begin{abstract}
In view of the great contribution of neutrino-electron scattering to the deep understanding of electroweak interactions, we focus in this paper on the study of elastic scattering of a muon neutrino by an electron $(e^{-}\nu_{\mu}\rightarrow e^{-}\nu_{\mu})$ in the presence of a circularly polarized electromagnetic field. We perform our theoretical calculation within the framework of Fermi theory using the exact wave functions of charged particles in an electromagnetic field. The expression of the differential cross section (DCS) for this process is obtained analytically in the absence and presence of the laser field. The effect of the field strength and frequency on the exchange of photons as well as on the DCS is presented and analyzed. Massive neutrino effects are also included and discussed. This study, added to the previous ones, will significantly enrich our knowledge in fundamental physics.
\end{abstract}

\maketitle
\section{Introduction}
In particle physics, the scattering experiments are the most effective research tools that allow us to study the interactions between particles and probe the structure of matter. It is through them that we gather a lot of information about the physical world. From the Rutherford's gold-foil experiment \cite{rutherford} revealing the atomic nucleus to the discovery of the Higgs boson at the Large Hadron Collider (LHC) \cite{higgs}, the observation and interpretation of quantum scattering processes have been pivotal to the advancement of particle physics. Laser-assisted or -induced scattering processes are a type of scattering processes that has received considerable attention since the invention of the laser in the $1960$s until the recent development of high-power optical lasers \cite{laser,laser1,laser2}. G. Mourou and D. Strickland were jointly awarded the 2018 Nobel Prize in Physics for inventing the use of chirped pulse amplification (CPA) as a means to generate high-intensity and ultra-short optical pulses \cite{nobel2018,mourou}. Meanwhile, scientists have been enthusiastic and focused on the study of laser-assisted quantum processes, whether in quantum electrodynamics (QED) \cite{qed0,hartin,qed1,qed2}, electroweak theory \cite{mouslih,jakha,muon1,muon2,muon3} or atomic physics \cite{atom,atom1,atom2,atom3}, and generally on the study of the laser-matter interactions \cite{bulanov,salamin}. In parallel with the production of the high-power femtosecond lasers, experimental testing of these studies became feasible \cite{test1,test2,test3}. In our turn, we have decided to study, through this research paper, one of the most important scattering processes that have played an important role in the development of modern physics, namely the elastic scattering process of a muon neutrino by an electron $(e^{-}\nu_{\mu}\rightarrow e^{-}\nu_{\mu})$ in the presence of a circularly polarized electromagnetic field. Nine years ago, the same process has been studied, within the framework of Fermi's theory, in the presence of a linearly polarized laser field and the phenomena of multi photon absorption and emission are induced \cite{bai2012}; the authors found that the distributions of a multiphoton energy transfer (MPET) spectrum are largely affected by the laser even at moderate intensities, while the differential cross  section (DCS) can be notably changed by several orders of magnitude only in superstrong fields with ultrarelativistic electrons. In general, the study of (anti)muon neutrino scattering on electrons has offered important results for understanding the electroweak sector of the Standard Model \cite{vilain}. In particular, the first observation of a few $\bar{\nu}_{\mu}e\rightarrow\bar{\nu}_{\mu}e$ scattering events by the Gargamelle bubble chamber in 1973 at CERN \cite{hasert} provided the first empirical proof of the existence of weak neutral-current (NC) interactions. An extensive theory and a collection of new results for total and differential cross sections of the elastic neutrino-electron scattering have been reported in Refs. \cite{tomalak,marciano}. The purpose of this paper is mainly to reveal the effect of a strong electromagnetic field on the scattering process $e^{-}\nu_{\mu}\rightarrow e^{-}\nu_{\mu}$ and in particular on its calculated DCS. The theoretical calculations are performed within the Fermi theory using the method of exact solutions for electron states in the presence of a circularly polarized electromagnetic wave field. Although the electromagnetic field does not couple to the standard model neutrino, it affects neutrino physics by altering the behavior of the charged particles with which the neutrino may interact \cite{tinsley,dicus}. Moreover, a short review of other processes involving neutrinos affected by the presence of a magnetic field can be found in Ref. \cite{pal}. In this paper, our plan is as follows. In order to proceed in a pedagogical way and to give the reader the necessary material, we present first, in Sec.~\ref{sec:free}, the detailed theoretical calculation of the DCS of the scattering process in the absence of any external field. Then in Sec.~\ref{sec:laser}, the explicit expression of the laser-assisted DCS is derived. The theoretical changes required when considering the nonzero neutrino mass are summarized in Sec.~\ref{sec:neutrinomass}. The results obtained, whether in the absence or in the presence of the laser, are presented and discussed in Sec.~\ref{sec:res}. Finally, Sec.~\ref{sec:conclusion} summarizes the results of this work and draws conclusions. The relativistic system of units, where $\hbar=c=1$, and the metric tensor $g^{\mu\nu}=diag(1, -1, -1, -1)$ will be used throughout this paper.
\section{Laser-free scattering process}\label{sec:free}
In this section, we calculate analytically the general formula of the unpolarized DCS for the scattering of a muon neutrino by an electron in the absence of a laser field. This scattering process can be schematized as
\begin{equation}\label{process}
e^{-}(p_{i})+\nu_{\mu}(k_{i})\longrightarrow e^{-}(p_{f})+\nu_{\mu}(k_{f}),
\end{equation}
where the arguments are our labels for the associated four-momentum for each particle, the indices $i$ and $f$ stand, respectively, for the initial and final states. This process is a weak interaction process; it can be described by the lowest Feynman diagrams. Therefore, in the first Born approximation, the transition matrix element can be written as
\begin{equation}\label{S-matrix}
S_{fi}=\dfrac{-iG}{\sqrt{2}}\int_{-\infty}^{\infty} d^{4}x\bigg[\overline{\psi}^{\textsl{f}}_{\nu_{\mu}}(x)\gamma^{\mu}(1-\gamma_{5})\psi^{\textsl{i}}_{\nu_{\mu}}(x)\bigg]\bigg[\overline{\psi}^{\textsl{f}}_{e^{-}}(x)\gamma_{\mu}(g_{V}-g_{A}\gamma_{5})\psi^{\textsl{i}}_{e^{-}}(x)\bigg],
\end{equation}
where $\textit{G}=(1.166~37\pm0.000~02)\times10^{-11}~\text{MeV}^{-2}$ is the coupling constant of Fermi measured from muon decay \cite{pdg2020}. $g_{V}=1/2-2\sin^{2}(\theta_{W})$ and $g_{A}=-1/2$ are, respectively, the vector and  axial-vector coupling constants and $ \theta_{W}$ is the weak-mixing angle.
Here, for $g_V$ and $g_A$, we choose the corrected (loop-level) values from the tree-level ones as follows: $g_{V}=0.043 \pm 0.063$ and $g_{A}=-0.545 \pm 0.056$ \cite{greiner}. The Dirac wave functions, normalized to the volume $V$, that describe the incoming and outgoing muon neutrinos have the following form:
\begin{equation}\label{wave function of neutrino}
\begin{split}
&\psi^{\textsl{i}}_{\nu_{\mu}}(x)=\dfrac{1}{\sqrt{2E_{i}V}}u_{\nu_{\mu}}(k_{i},t_{i})e^{-ik_{i}.x},\\
&\psi^{\textsl{f}}_{\nu_{\mu}}(x)=\dfrac{1}{\sqrt{2E_{f}V}}u_{\nu_{\mu}}(k_{f},t_{f})e^{-ik_{f}.x},
\end{split}
\end{equation}
where $u_{\nu_{\mu}}(k_{i,f},t_{i,f})$ represents the Dirac bispinor with four-momentum $k_{i,f}$ and spin $t_{i,f}$ satisfying 
\begin{equation}\label{nuspinor}
\sum_{t_{i,f}}\bar{u}_{\nu_{\mu}}(k_{i,f},t_{i,f})u_{\nu_{\mu}}(k_{i,f},t_{i,f})= \slashed{k}_{i,f}.
\end{equation}
In the first stage, we will consider the muon neutrinos as massless particles. The  wave functions $\psi_{e^{-}}^{\textsl{i}}(x)$ and $\psi_{e^{-}}^{\textsl{f}}(x)$ are, respectively, the initial and final states of the electron
\begin{equation}\label{wave function of electron}
\begin{split}
&\psi^{\textsl{i}}_{e^{-}}(x)=\dfrac{1}{\sqrt{2p_{i}^{0}V}}u_{e^{-}}(p_{i},s_{i})e^{-ip_{i}.x},\\
&\psi^{\textsl{f}}_{e^{-}}(x)=\dfrac{1}{\sqrt{2p_{f}^{0}V}}u_{e^{-}}(p_{f},s_{f})e^{-ip_{f}.x},
\end{split}
\end{equation}
where $p_{i}^{0}$ and $p_{f}^{0}$  are, respectively, the total energies of the incident and outgoing electrons. $u_{e^{-}}(p_{i,f},s_{i,f})$ is the free Dirac spinor satisfying $\sum_{s_{i,f}}\bar{u}_{e^{-}}(p_{i,f},s_{i,f})u_{e^{-}}(p_{i,f},s_{i,f})= \slashed{p}_{i,f}+m $,  where $m$ is the rest mass of the electron. After introducing the different wave functions describing the particles involved in the process (\ref{process}), we substitute them into the S-matrix element (\ref{S-matrix}). After some manipulations, we get
\begin{equation}
S_{fi}=\dfrac{-i G}{\sqrt{32E_{i}E_{f}p_{i}^{0}p_{f}^{0} V^{4}}} (2\pi)^{4}\delta^{4}(p_{f}+k_{f}-p_{i}-k_{i})M_{fi},
\end{equation}
where
\begin{equation}\label{mfi_free}
M_{fi}= \big[\bar{u}_{\nu_{\mu}}(k_{f},t_{f})\gamma^{\mu}(1-\gamma^{5})u_{\nu_{\mu}}(k_{i},t_{i}) \big]\big[\bar{u}_{e^{-}}(p_{f},s_{f})\gamma_{\mu}(g_{V}-g_{A}\gamma_{5})u_{e^{-}}(p_{i},s_{i}) \big].
\end{equation}
What interests us in the evaluation of the DCS is the square of the S-matrix element $|S_{fi}|^{2}$ multiplied by the density of final states and divided by the flux of the incoming particles $|J_{inc}|$. In the case of unpolarized DCS, we must sum over the final spin states and average over the initial spin states. We point out here, by the way, that electrons can be in two spin states, while neutrinos exist in only one state of negative helicity \cite{greiner}. This yields
\begin{equation}
\begin{split}
d\overline{\sigma}(e^{-} \nu_{\mu} \rightarrow e^{-} \nu_{\mu})&=~V\int\dfrac{d^{3}p_{f}}{(2\pi)^{3}}~V\int\dfrac{d^{3}k_{f}}{(2\pi)^{3}}~\dfrac{1}{2}\sum_{t_{i,f},s_{i,f}}\dfrac{|S_{fi}(e^{-} \nu_{\mu} \rightarrow e^{-} \nu_{\mu})|^{2}}{ T|J_{inc}|},\\
&=\dfrac{G^{2}}{64p_{i}^{0}p_{f}^{0}E_{i}E_{f} V}\dfrac{\delta^{4}(p_{f}+k_{f}-p_{i}-k_{i})}{(2\pi)^{2}|J_{inc}|}\int d^{3}k_{f}\int d^{3}p_{f}\sum_{t_{i,f},s_{i,f}}|M_{fi}|^{2},
\end{split}
\end{equation}
where we have used $[(2\pi)^{4}\delta^{4}(p_{f}+k_{f}-p_{i}-k_{i})]^{2}= V T (2\pi)^{4}\delta^{4}(p_{f}+k_{f}-p_{i}-k_{i})$ and $ J_{inc}$ is the incoming neutrinos current in the laboratory system given by $J_{inc}=(k_{i}.p_{i})/(E_{i} p_{i}^{0} V )$. With the help of the following relations $d^{3}p_{f}=|\textbf{p}_{f}|^{2}d|\textbf{p}_{f}|d\Omega $ and $\delta^{4}(p_{f}+k_{f}-p_{i}-k_{i})=\delta^{0}(p_{f}^{0}+E_{f}-p_{i}^{0}-E_{i})\delta^{3}(\textbf{p}_{f}+\textbf{k}_{f}-\textbf{p}_{i}-\textbf{k}_{i})$, the DCS  becomes
\begin{equation}
\dfrac{d\overline{\sigma}}{d\Omega}(e^{-} \nu_{\mu} \rightarrow e^{-} \nu_{\mu})=\dfrac{G^{2}}{64(2\pi)^{2}(k_{i}.p_{i})}\int\dfrac{|\textbf{p}_{f}|^{2}d|\textbf{p}_{f}|}{E_{f}p_{f}^{0}} \delta^{0}(p_{f}^{0}+E_{f}-p_{i}^{0}-E_{i})\sum_{t_{i,f},s_{i,f}}|M_{fi}|^{2}\bigg|_{\textbf{p}_{f}+\textbf{k}_{f}-\textbf{p}_{i}-\textbf{k}_{i}=0.}
\end{equation}
The remaining integral over $ d|\textbf{p}_{f}|$ can be solved by using the following formula \cite{greiner} 
\begin{align}\label{familiarformula}
\int dxf(x)\delta(g(x))=\dfrac{f(x)}{|g'(x)|}\bigg|_{g(x)=0.}
\end{align}
Finally, we get
\begin{equation}\label{freedcs}
\begin{split}
\dfrac{d\overline{\sigma}}{d\Omega}(e^{-} \nu_{\mu} \rightarrow e^{-} \nu_{\mu})=&\dfrac{G^{2}}{ 256 \pi^{2} p_{f}^{0}E_{f}}\dfrac{|\textbf{p}_{f}|^{2}}{(k_{i}.p_{i}) |g'(|\textbf{p}_{f}|)|}\\& \times \text{Tr}\big[(\slashed{p}_{f}+m)\gamma_{\mu}(g_{V}-g_{A}\gamma_{5})(\slashed{p}_{i}+m)\gamma_{\nu}(g_{V}-g_{A}\gamma_{5})\big]\\& \times\text{Tr}\big[\slashed{k}_{f}\gamma^{\mu}(1-\gamma^{5})\slashed{k}_{i}\gamma^{\nu}(1-\gamma^{5})\big],
\end{split}
\end{equation}
where
\begin{equation}
g'(|\textbf{p}_{f}|)=\dfrac{|\textbf{p}_{f}|}{\sqrt{|\textbf{p}_{f}|^{2}+m^{2}}}+\dfrac{|\textbf{p}_{f}|+E_{i}\cos(\theta_{f})-|\textbf{p}_{i}|F(\phi_{i},\phi_{f},\theta_{i},\theta_{f})}{E_{f}},
\end{equation}
and 
\begin{equation}
\begin{split}
F(\phi_{i},\phi_{f},\theta_{i},\theta_{f})=&\cos(\phi_{i})\sin(\theta_{i})\cos(\phi_{f})\sin(\theta_{f})+\sin(\theta_{i})\sin(\phi_{i})\sin(\theta_{f})\sin(\phi_{f})\\
&+\cos(\theta_{i})\cos(\theta_{f}).
\end{split}
\end{equation}
The product of the two traces in Eq.~(\ref{freedcs}) is evaluated and gives the following result:
\begin{equation}\label{restracefree}
\begin{split}
\sum_{t_{i,f},s_{i,f}}|M_{fi}|^{2}=& 64 \big[g_A^2 \big((k_f.p_f)(k_i.p_i)+(k_f.p_i)(k_i.p_f)+(k_i.k_f)m^2\big)+2g_A g_V \big((k_f.p_f)(k_i.p_i)\\&-(k_f.p_i)(k_i.p_f)\big)+g_V^2 \big((k_f.p_f)(k_i.p_i)+(k_f.p_i)(k_i.p_f)-(k_i.k_f)m^2\big)\big].
\end{split}
\end{equation}
The reader can refer to the textbook in Ref.~\cite{greiner} for another independent, pedagogical, and more detailed calculation of this scattering process in the absence of a laser field.
\section{Laser-assisted scattering process}\label{sec:laser}
Now, we consider the process (\ref{process}) in the presence of a laser field. For reasons of mathematical simplicity, the laser field is considered as a monochromatic plane wave of circular polarization, whose classical four-potential can be expressed in a unified notation such as
\begin{equation}
A^{\mu}(x)=|\textbf{a}| \big[\eta_{1}^{\mu} \cos(k.x)+ \eta_{2}^{\mu} \sin(k.x)\big],
\end{equation}
where $|\textbf{a}|=\mathcal{E}_{0}/\omega$ denotes the magnitude of the four-potential with $\mathcal{E}_{0}$ is the electric field strength and $\omega$ is the laser frequency. $k=(\omega,0,0,\omega)$ is the wave 4-vector chosen to be directed along the $z$-axis. The polarization 4-vectors $\eta_{1}^{\mu}$ and $ \eta_{2}^{\mu}$ satisfy the normalization and orthogonality conditions, $\eta_{1}^{2}=\eta_{2}^{2}=-1$ and $(\eta_{1}.\eta_{2})=0$. The Lorentz gauge condition, $k_{\mu}.A^{\mu}=0$, applied to the four-potential implies $(k.\eta_{1})=0$ and  $(k.\eta_{2})=0$. Now, in the presence of a laser field, the electron obeys the following Dirac equation \cite{landau}:
\begin{equation}
\big[(p_{i,f}-e A)^{2}-m^{2}-\dfrac{i e}{2}  F_{\mu \nu} \sigma^{\mu \nu}\big]\psi^{\textsl{i,f}}_{e^{-}}(x)=0,
\end{equation}
where $e=-|e|<0$ is the charge of the electron, $F_{\mu \nu}=\partial_{\mu} A_{\nu}-\partial_{\nu} A_{\mu}$ is the electromagnetic field tensor and $\sigma^{\mu \nu}=\tfrac{1}{2}[\gamma^{\mu},\gamma^{\nu}]$. The general solution of this equation is the normalized relativistic Dirac-Volkov wave functions \cite{volkov}, which describe the incident and outgoing electrons in a laser field 
\begin{equation}\label{ewavefunction}
\psi^{\textsl{i,f}}_{e^{-}}(x)=\bigg[1+\dfrac{e\slashed{k}\slashed{A}}{2(k.p_{i,f})}\bigg]\frac{u(p_{i,f},s_{i,f})}{\sqrt{2Q_{i,f}V}}\times e^{iS(q_{i,f},x)},
\end{equation}
where 
\begin{equation}
S(q_{i,f},x)=-q_{i,f}.x-\dfrac{e|\textbf{a}|(\eta_{1}.p_{i,f})}{(k.p_{i,f})}\sin(k.x)+\dfrac{e|\textbf{a}|(\eta_{2}.p_{i,f})}{(k.p_{i,f})}\cos(k.x).
\end{equation}
$ q_{i,f}=(Q_{i,f},\textbf{q}_{i,f})$ is the Volkov momentum of the electron in the presence of the laser field, which is given by
\begin{equation}
q_{i,f}=p_{i,f}+\frac{e^{2}|\textbf{a}|^{2}}{2(k.p_{i,f})}k.
\end{equation}
Squaring this four-momentum shows that the mass of the dressed electron (effective mass) has a dependence on the strength of the EM field
\begin{equation}
m_{*}^{2}=m^{2}+e^{2}|\textbf{a}|^{2}.
\end{equation}
Note that in the absence of the EM field $(|\textbf{a}|\rightarrow 0)$, the Volkov wave function, Eq.~(\ref{ewavefunction}), reduces to the free-field wave function given in Eq.~(\ref{wave function of electron}), and the mass $m_{*}$ and four-momentum $q_{i,f}$ of the dressed electron reduce also in that case to the electron mass $m$ and four-momentum $p_{i,f}$, respectively.
Since the incoming and outgoing muon neutrinos are electrically neutral and thus do not interact with the laser field, their wave functions are unchanged. We will therefore describe them using the same previous unaffected plane waves given in Eq.~(\ref{wave function of neutrino}). Thus, the transition matrix element  (\ref{S-matrix}) becomes
\begin{equation}
\begin{split}
S_{fi}=&\dfrac{-i G}{\sqrt{32E_{i}E_{f}Q_{i}Q_{f} V^{4}}}\int d^{4}x~e^{i(k_{f}-k_{i}).x}~e^{i(S(q_{i},x)-S(q_{f},x))}
\Big[\bar{u}(p_{f},s_{f})\Big(1+C(p_{f})\slashed{A}\slashed{k}\Big)\\
&\times\gamma_{\mu}(g_{V}-g_{A}\gamma_{5})\Big(1+C(p_{i})\slashed{k}\slashed{A}\Big)u(p_{i},s_{i}) \Big]\Big[\bar{u}_{\nu_{\mu}}(k_{f},t_{f})\gamma^{\mu}(1-\gamma^{5})u_{\nu_{\mu}}(k_{i},t_{i}) \Big],
\end{split}
\end{equation}
where $C(p_{i})=e/(2(k.p_{i}))$ and $C(p_{f})=e/(2(k.p_{f}))$. We can recast the exponential term in the appropriate form, $e^{i(S(q,x)-S(q_{1},x))}=e^{i(q_{f}-q_{i}).x} ~ e^{-iz\sin(k.x-\varphi)}$, where
\begin{equation}\label{argument}
z=e |\textbf{a}|\sqrt{\bigg(\dfrac{\eta_{1}.p_{i}}{k.p_{i}}-\dfrac{\eta_{1}.p_{f}}{k.p_{f}}\bigg)^{2}+\bigg(\dfrac{\eta_{2}.p_{i}}{k.p_{i}}-\dfrac{\eta_{2}.p_{f}}{k.p_{f}}\bigg)^{2}},
\end{equation}
and
\begin{equation}
\varphi=\arctan\bigg[\frac{(\eta_{2}.p_{i})(k.p_{f})-(\eta_{2}.p_{f})(k.p_{i})}{(\eta_{1}.p_{i})(k.p_{f})-(\eta_{1}.p_{f})(k.p_{i})}\bigg].
\end{equation}
Therefore, the transition  matrix element becomes
\begin{equation}\label{s_matrix element}
\begin{split}
S_{fi}=&\dfrac{-i G}{\sqrt{32E_{i}E_{f}Q_{i}Q_{f} V^{4}}}\int d^{4}x~e^{i(k_{f}+q_{f}-k_{i}-q_{i}).x}~e^{-iz\sin(k.x-\varphi)}
\Big[\bar{u}(p_{f},s_{f})\big(\Delta_{0\mu}\\
&+\Delta_{1\mu}\cos(k.x)+\Delta_{2\mu}\sin(k.x)\big)u(p_{i},s_{i}) \Big]\Big[\bar{u}_{\nu_{\mu}}(k_{f},t_{f})\gamma^{\mu}(1-\gamma^{5})u_{\nu_{\mu}}(k_{i},t_{i}) \Big],
\end{split}
\end{equation}
where
\begin{equation}\label{Deltas}
\begin{split}
&\Delta_{0\mu}=\gamma_{\mu}~(g_{V}-g_{A}\gamma_{5})+2~ C(p_{i})~ C(p_{f})~ |\textbf{a}|^{2}~k_{\mu}~\slashed{k}~(g_{V}-g_{A}\gamma_{5}),\\
& \Delta_{1\mu}=C(p_{i})|\textbf{a}| \gamma_{\mu}~(g_{V}-g_{A}\gamma_{5}) \slashed{k}~\slashed{\eta}_{1}+C(p_{f})|\textbf{a}|~\slashed{\eta}_{1}~\slashed{k}~\gamma_{\mu}~(g_{V}-g_{A}\gamma_{5}),\\
& \Delta_{2\mu}=C(p_{i})|\textbf{a}| \gamma_{\mu}~(g_{V}-g_{A}\gamma_{5}) \slashed{k}~\slashed{\eta}_{2}+C(p_{f})|\textbf{a}|~\slashed{\eta}_{2}~\slashed{k}~\gamma_{\mu}~(g_{V}-g_{A}\gamma_{5}).
\end{split}
\end{equation}
The linear combination of the three different quantities in Eq.~(\ref{s_matrix element}) can be transformed by the well-known Jacobi-Anger identity involving ordinary Bessel functions $J_{n}(z)$
\begin{equation}\label{transformation}
\begin{split}
\begin{Bmatrix}
1\\
\cos(k.x)\\
\sin(k.x)
\end{Bmatrix}\times e^{-iz\sin(k.x-\varphi)}&=\sum_{n=-\infty}^{+\infty}e^{-in(k.x)}\begin{Bmatrix}
J_{n}(z)e^{in\varphi}\\
\frac{1}{2}\big\lbrace J_{n+1}(z)e^{i(n+1)\varphi}+J_{n-1}(z)e^{i(n-1)\varphi}\big\rbrace\\
\frac{1}{2i}\big\lbrace J_{n+1}(z)e^{i(n+1)\varphi}-J_{n-1}(z)e^{i(n-1)\varphi}\big\rbrace
\end{Bmatrix},\\&
=\sum_{n=-\infty}^{+\infty}e^{-in(k.x)}\begin{Bmatrix}
b_{n}(z)\\
b_{1n}(z)\\
b_{2n}(z) \end{Bmatrix},
\end{split}
\end{equation}
where $z$ is the argument of the Bessel functions defined in Eq.~(\ref{argument}) and $n$, their order, is interpreted as the number of exchanged photons. After the integration, the S-matrix element $S_{fi}$  becomes
\begin{equation}
\begin{split}
S_{fi}=\dfrac{-i G}{\sqrt{32E_{i}E_{f}Q_{i}Q_{f} V^{4}}}\sum_{n=-\infty}^{+\infty}(2\pi)^{4}\delta^{4}(q_{f}+k_{f}-q_{i}-k_{i}-n k)~M^{n}_{fi}.
\end{split}
\end{equation}
The quantity $M^{n}_{fi}$ is defined by
\begin{equation}\label{mfin}
M^{n}_{fi}=\big[\bar{u}(p_{f},s_{f})\Gamma_{\mu}^{n}u(p_{i},s_{i}) \big]\big[\bar{u}_{\nu_{\mu}}(k_{f},t_{f})\gamma^{\mu}(1-\gamma^{5})u_{\nu_{\mu}}(k_{i},t_{i}) \big],
\end{equation}
where
\begin{equation}\label{Gamma_mu}
\Gamma_{\mu}^{n}= \Delta_{0\mu} b_{n}(z)+\Delta_{1\mu} b_{1n}(z)+\Delta_{2\mu} b_{2n}(z).
\end{equation}
To evaluate the DCS in the presence of a laser field, we follow the same steps as in the absence of the laser field in the previous section. This yields
\begin{equation}
\begin{split}
d\overline{\sigma}=&\sum_{n,l=-\infty}^{+\infty}\dfrac{G^{2}}{32E_{i}E_{f}Q_{i}Q_{f}V^4T|J_{inc}|} V\int\dfrac{d^{3}q_{f}}{(2\pi)^{3}}~V\int\dfrac{d^{3}k_{f}}{(2\pi)^{3}}~\dfrac{1}{2}\sum_{t_{i,f},s_{i,f}} (2\pi)^{8}\delta^{4}(q_{f}+k_{f}-q_{i}-k_{i}-n k)\\&\times \delta^{4}(q_{f}+k_{f}-q_{i}-k_{i}-l k) M^{l*}_{fi}M^{n}_{fi}.\\
\end{split}
\end{equation}
From the inspection, one can see that both 4-dimensional $\delta$ functions imply that for there to be any contribution to the summation, either there is no incoming photon energy ($E_{\gamma}=0$) or, more appropriately, that $l=n$. We can therefore replace $M^{l*}_{fi}M^{n}_{fi}$ by the square of the norm of the scattering amplitude $|M^{n}_{fi}|^2$ and exclude the sum over $l$. We get for the unpolarized  DCS
\begin{equation}
\begin{split}
d\overline{\sigma}=&\sum_{n=-\infty}^{+\infty}\dfrac{G^{2}}{64E_{i}E_{f}Q_{i}Q_{f}V^2T|J_{inc}|} \int\dfrac{d^{3}q_{f}}{(2\pi)^{3}}~\int\dfrac{d^{3}k_{f}}{(2\pi)^{3}} [(2\pi)^{4}\delta^{4}(q_{f}+k_{f}-q_{i}-k_{i}-n k)]^2 \sum_{t_{i,f},s_{i,f}} |M^{n}_{fi}|^2,\\
\end{split}
\end{equation}
which is simplified after some manipulation into the following form:
\begin{equation}
\dfrac{d\overline{\sigma}}{d\Omega}=\sum_{n=-\infty}^{+\infty}\dfrac{G^{2}}{64Q_{f}Q_{i}E_{f}E_{i}}\dfrac{|\textbf{q}_{f}|^{2}d|\textbf{q}_{f}|}{(2\pi)^{2}|J_{inc}|V} \delta^{0}(Q_{f}+E_{f}-Q_{i}-E_{i}-n\omega)\sum_{t_{i,f},s_{i,f}}|M^{n}_{fi}|^{2}\bigg|_{\textbf{q}_{f}+\textbf{k}_{f}-\textbf{q}_{i}-\textbf{k}_{i}-n\textbf{k}=0.}
\end{equation}
Using the formula (\ref{familiarformula}) to perform the remaining integral over $d|\textbf{q}_{f}|$ and replacing the incoming current $|J_{inc}|=(k_{i}.q_{i})/(Q_{i}E_{i}V) $, we obtain 
\begin{equation}\label{dcswithlaser}
\begin{split}
\Big(\dfrac{d\overline{\sigma}}{d\Omega}\Big)^{\text{with laser}}=\sum_{n=-\infty}^{+\infty}\dfrac{d\overline{\sigma}^{n}}{d\Omega}=&\sum_{n=-\infty}^{+\infty}\dfrac{G^{2}}{256\pi^{2}Q_{f}E_{f}}\dfrac{|\textbf{q}_{f}|^{2}}{(k_{i}.q_{i})|g'(|\textbf{q}_{f}|)|}\\& \times \text{Tr}\big[(\slashed{p}_{f}+m)\Gamma_{\mu}^{n}(\slashed{p}_{i}+m)\overline{\Gamma}_{\nu}^{n}\big]\\&\times \text{Tr}\big[\slashed{k}_{f}\gamma^{\mu}(1-\gamma^{5})\slashed{k}_{i}\gamma^{\nu}(1-\gamma^{5})\big],
\end{split}
\end{equation}
where
\begin{equation}
g'(|\textbf{q}_{f}|)= \dfrac{|\textbf{q}_{f}|}{\sqrt{|\textbf{q}_{f}|^{2}+m_{*}^{2}}}+\dfrac{|\textbf{q}_{f}|+E_{i}\cos(\theta_{f})-n\omega\cos(\theta_{f})-|\textbf{q}_{i}|F(\phi_{i},\phi_{f},\theta_{i},\theta_{f})}{E_{f}},
\end{equation}
and
\begin{equation}
\begin{split}
\overline{\Gamma}_{\nu}^{n}&=\gamma^{0} \Gamma_{\nu}^{n+}\gamma^{0},\\
&= \overline{\Delta}_{0\nu}b_{n}^{*}(z)+ \overline{\Delta}_{1\nu}b_{1n}^{*}(z)+ \overline{\Delta}_{2\nu}b_{2n}^{*}(z),
\end{split}
\end{equation}
where
\begin{equation}
\begin{split}
&\overline{\Delta}_{0\nu}=\gamma_{\nu}~(g_{V}-g_{A}\gamma_{5})+2~ C(p_{i})~ C(p_{f})~ |\textbf{a}|^{2}~k_{\nu}~\slashed{k}~(g_{V}-g_{A}\gamma_{5}),\\
& \overline{\Delta}_{1\nu}=C(p_{i}) |\textbf{a}|~\slashed{\eta}_{1}~\slashed{k}\gamma_{\nu}~(g_{V}-g_{A}\gamma_{5}) +C(p_{f})|\textbf{a}|~\gamma_{\nu}~(g_{V}-g_{A}\gamma_{5})\slashed{k}~\slashed{\eta}_{1},\\
&\overline{\Delta}_{2\nu}=C(p_{i}) |\textbf{a}|~\slashed{\eta}_{2}~\slashed{k}\gamma_{\nu}~(g_{V}-g_{A}\gamma_{5}) +C(p_{f})|\textbf{a}|~\gamma_{\nu}~(g_{V}-g_{A}\gamma_{5})\slashed{k}~\slashed{\eta}_{2}.
\end{split}
\end{equation}
The evaluation of the traces is commonly performed with the help of FEYNCALC \cite{feyncalc1,feyncalc2,feyncalc3}. The result we obtained is attached in the Appendix. In order to distinguish between them and to avoid any confusion, it would be very appropriate to consider the summed differential cross section (SDCS), $(d\overline{\sigma}/d\Omega)^{\text{with laser}}$, as the sum of discrete individual differential cross sections (IDCS), $d\overline{\sigma}^{n}/d\Omega$, for each photon exchange process.\\
Note that in the absence of an EM field (i.e., when we take the limit $|\textbf{a}|\rightarrow 0$ and without the exchange of any photons $(n=0)$), the argument of the Bessel functions vanishes $(z=0)$. Thus, all the terms that contribute to $\Gamma_{\mu}^{n}$ given in Eq.~(\ref{Gamma_mu}) will be null except for the term $\Delta_{0\mu}$, since $b_{0}(0)=1$ and $b_{10}(0)=b_{20}(0)=0$. The expression of $\Delta_{0\mu}$ in Eq.~(\ref{Deltas}) reduces, for $|\textbf{a}|\rightarrow 0$, to $\gamma_{\mu}~(g_{V}-g_{A}\gamma_{5})$, and then the quantity $M^{n=0}_{fi}$ (in Eq.~(\ref{mfin})) becomes that obtained in the absence of the laser field (Eq.~(\ref{mfi_free})). The same thing applies to the S-matrix element $S_{fi}$ and all 
other quantities that compose the DCS in Eq.~(\ref{dcswithlaser}).
\section{Nonzero neutrino mass effect}\label{sec:neutrinomass}
Throughout the calculation performed in the two previous sections, we have disregarded the mass of the muon neutrinos $(m_{\nu_{\mu}}=0)$. In this section, and in order to reveal the effect induced by the nonzero mass of neutrinos, we will show what changes will theoretically occur in some of the relations when considering the mass of neutrinos. But first, we will try to give some brief words about the discovery that neutrinos have a nonnegligible mass. Within the standard model of elementary particle physics, the masses of the neutrinos are assumed to be exactly zero. It is only recently that it has been confirmed that neutrinos have small but nonzero masses by discovering the phenomenon of neutrino oscillations \cite{kajita2010}. Neutrino oscillation is a quantum mechanical phenomenon in which a neutrino produced in a specific lepton flavor can later be measured to have a different lepton flavor after traveling some distances. Neutrino oscillation has been discovered through studies of neutrinos produced by cosmic-ray interactions in the atmosphere \cite{kajita2006,fukuda98}. Since we are interested in the present research on the muon neutrino $\nu_{\mu}$,
its discovery in 1962 \cite{danby62} by Leon Lederman, Melvin Schwartz and Jack Steinberger was rewarded with the 1988 Nobel Prize in Physics \cite{nobel1988}. An evidence of its oscillations to $\nu_{\tau}$ in the KEK to Kamioka (K2K) accelerator-based experiment has been reported in Ref. \cite{aliu2005}. An upper limit of 0.17 MeV (at the $90\%$ confidence level) for the muon-neutrino mass $(m_{\nu_{\mu}}<0.17~\text{MeV})$ has been derived in \cite{assamagan} by  analyzing the decay at rest of positively charged pions. To be more precise, it is very important to point out here that in spite of all the above, the mass of the muon neutrino remains a not well defined principle, and in contrast we introduce what is called the effective muon neutrino mass, which is an incoherent sum of neutrino masses and lepton mixing matrix elements \cite{katrin}. For illustration purposes and in order to make the effect visible we will take some liberties in the determination of the muon neutrino mass.
On this basis, when presenting the results for the nonzero neutrino mass, we choose the muon neutrino mass as $m_{\nu_{\mu}}=0.15~\text{MeV}$.
Now, let us summarize some of the changes that will occur in the theoretical calculation.
We start with the relation, Eq.~(\ref{nuspinor}), that Dirac bispinor of neutrino verifies, which will be the following: $\sum_{t_{i,f}}\bar{u}_{\nu_{\mu}}(k_{i,f},t_{i,f})u_{\nu_{\mu}}(k_{i,f},t_{i,f})= \slashed{k}_{i,f}+m_{\nu_{\mu}}.$ This change will be manifested in the calculation of the trace related to the neutrino current. Note also that the relation $E_{i,f}=|\textbf{k}_{i,f}|$ will not be valid anymore, we retain instead the original relativistic energy-momentum relation, $E_{i,f}=\sqrt{|\textbf{k}_{i,f}|^{2}+m_{\nu_{\mu}}^{2}}$. The other thing that will change is the incoming neutrinos flux $J_{inc}$, which will remain in its original formula as follows: $J_{inc}=\sqrt{(k_{i}.p_{i})^{2}-m_{\nu_{\mu}}^{2}m^{2}}/(E_{i} p_{i}^{0} V )$. With all of this, the entire dependence on the mass of neutrino is fully taken into account. The results of this effect on the DCS will be presented and discussed in the next section.
\section{Results and discussion}\label{sec:res}
This section will be devoted to the presentation and discussion of numerical results related to the DCS of the electron and muon neutrino elastic scattering process (\ref{process}) in the absence and presence of the electromagnetic field. The DCS is derived with respect to the solid angle of the electron due to the fact that neutrinos are somewhat difficult to capture experimentally. In order to obtain the total cross section (TCS), the integral over that solid angle must be computed using numerical methods. For the geometry, we consider the momentum of the incoming muon neutrino $\textbf{k}_{i}$, with kinetic energy $E_{\nu}^{\text{kin}}$, along the opposite direction of the $z$-axis. We set both the initial and final electrons in a general geometry with spherical coordinates $\theta_{i}$, $\theta_{f}$, $\phi_{i}$ and $\phi_{f}$. The angles $\phi_{i}$ and $\phi_{f}$ were chosen to be zero ($\phi_{i}=\phi_{f}=0^{\circ}$) in all results presented in this section. The momentum of the outgoing muon neutrino $\textbf{k}_{f}$ can be deduced from the previous ones by using the momentum conservation relationship, $\textbf{q}_{f}+\textbf{k}_{f}-\textbf{q}_{i}-\textbf{k}_{i}-n\textbf{k}=0$. Considering the effects of relativity and spin, we choose the kinetic energy of the incoming electron as (unless otherwise stated) $E_{e}^{\text{kin}}=10^{6}~\text{eV}$, and that of muon neutrinos varies in the range between $0.1$ and $10^{15}~\text{eV}$ depending on their sources \cite{formaggio2013}. 
\subsection{Without laser field}
\begin{figure}[hptb]
 \centering
\includegraphics[scale=0.5]{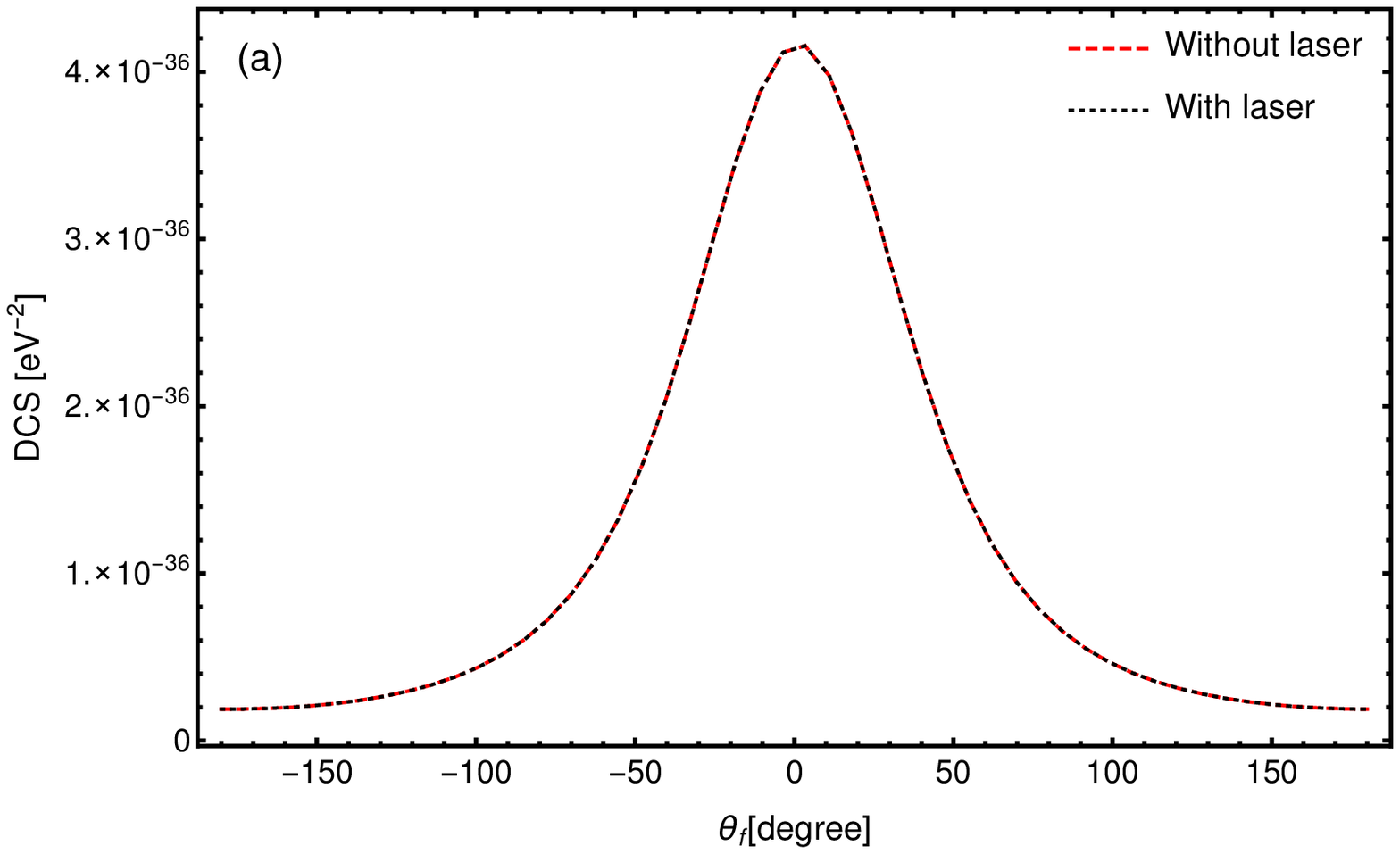}\hspace*{0.4cm}
\includegraphics[scale=0.505]{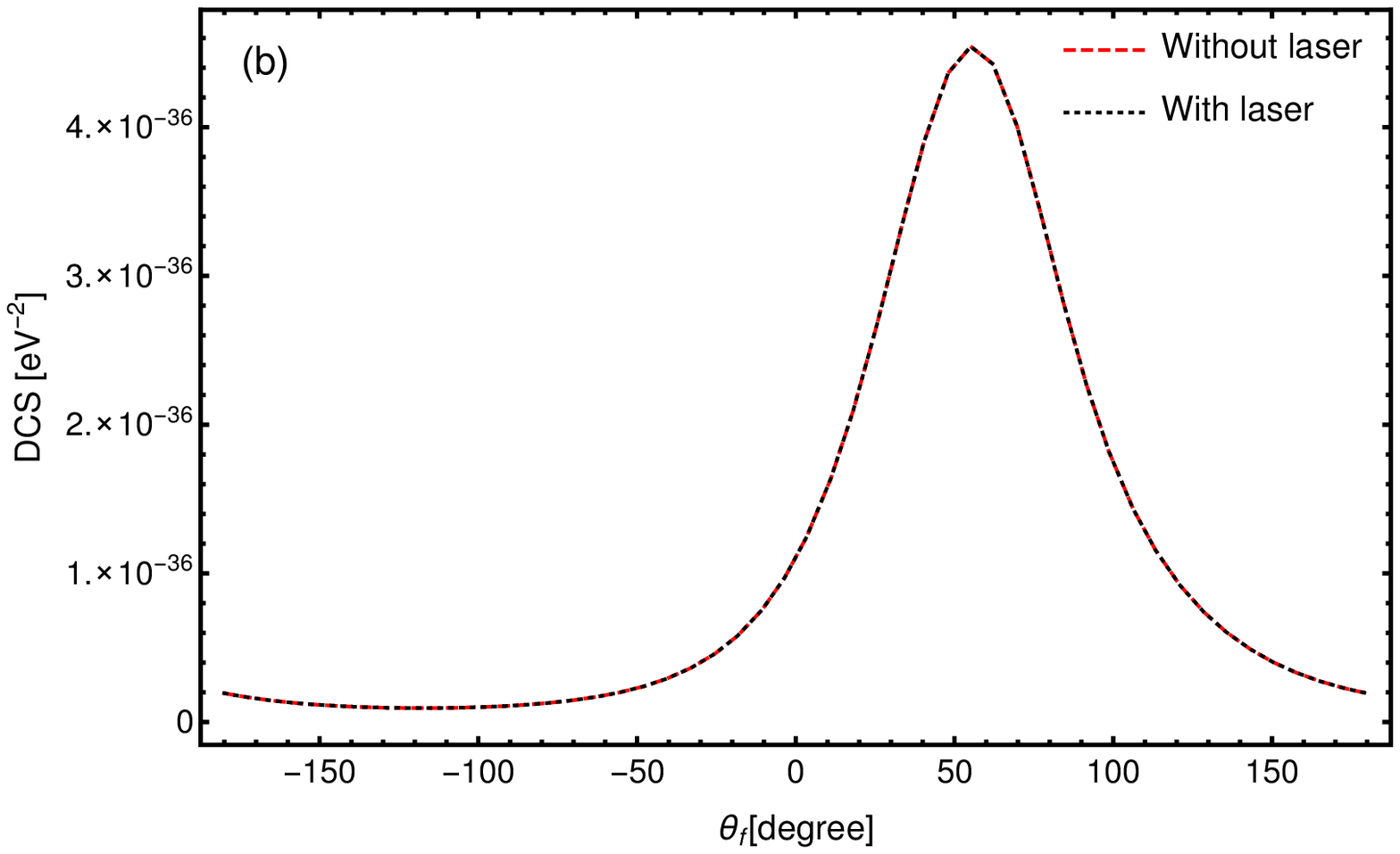}
\caption{The two DCSs with and without laser, drawn as a function of the scattering angle $\theta_{f}$ for an electric field strength of $\mathcal{E}_{0}=0~\text{V/cm}$ and without any exchange of photons ($n=0$). The kinetic energy of the incoming muon neutrino is $E_{\nu}^{\text{kin}}=0.5\times10^{6}$ eV. The initial angle $\theta_{i}$ is (a) $\theta_{i}=1^{\circ}$ and (b) $\theta_{i}=45^{\circ}$.}\label{fig1}
\end{figure}

In addition to pedagogical purposes, the main goal of performing calculations in the absence of a laser field is simply to provide a means that enables us to ensure the consistency of our calculations when a laser is added. This is done by taking the limit of the field strength $\mathcal{E}_{0}$ and the number of photons exchanged $n$ as they all tend to zero and then checking if the dressed DCS gives, under these conditions, its corresponding one in the absence of the laser. By the way, this check has been carried out successfully and the result is shown in Fig.~\ref{fig1}. 
Due to their perfect overlap, it is hardly possible to distinguish between the two curves that represent the changes in the two DCSs (with and without laser) as a function of the final angle $\theta_{f}$ of the electron. This represents a consistency check of our calculations. The difference between Figs.~\ref{fig1}\textcolor{blue}{(a)} and \ref{fig1}\textcolor{blue}{(b)} is only in the geometry, where in Fig.~\ref{fig1}\textcolor{blue}{(a)} the initial angle is $\theta_{i}=1^{\circ}$ and in Fig.~\ref{fig1}\textcolor{blue}{(b)} it is $\theta_{i}=45^{\circ}$. We note here that the choice $\theta_{i}=1^{\circ}$ was intentionally made only to avoid the divergence that occurs in the DCS in the case of $\theta_{i}=\theta_{f}=0^{\circ}$. In Fig.~\ref{fig1}\textcolor{blue}{(a)}, the DCS exhibits a symmetry in its curve with respect to the angle $\theta_{f}=0^{\circ}$, where it has a maximum value, in contrast to Fig.~\ref{fig1}\textcolor{blue}{(b)} in which the DCS is asymmetrical and its peak is shifted to large final angles ($\theta_{f}\simeq50^{\circ}$). This clearly shows the effect of geometry on the angular distribution of the DCS. Another thing worth paying attention to when examining these figures is to get information about the final angle $\theta_{f}$ at which the electron will go out after scattering from the initial angle $\theta_{i}$ at which it comes in. On the basis of these two figures, it appears that the peak is located approximately around the final angle that is equal to the initial one ($\theta_{f}\sim \theta_{i}$). This fact is valid for any geometry, provided that $\phi_{i}=\phi_{f}=0^{\circ}$. This means that an electron incoming at a small angle $\theta_{i}$ has a high probability to be scattered at a small final angle $\theta_{f}$ with a value approximately equal to the initial angle. As long as $\theta_{i}$ increases, the final angle $\theta_{f}$ also increases.\\
For further comparison, we give here a value for the total cross section obtained theoretically at the kinetic energy of the electron $E_{e}^{\text{kin}}=0.05\times10^{6}$ eV:
\begin{equation}\label{tcsvalueth}
\frac{1}{E_{\nu}^{\text{kin}}}\overline{\sigma}(e^{-}\nu_{\mu}\rightarrow e^{-}\nu_{\mu})=1.40\times 10^{-42}~\text{cm}^{2}/\text{GeV},
\end{equation}
where $E_{\nu}^{\text{kin}}=0.5\times10^{6}$ eV, and the error from the numerical integration performed using Simpson's rule is estimated to be less than $10^{-18}$.
This theoretical value given in Eq.~(\ref{tcsvalueth}) is to be compared with that obtained experimentally \cite{greiner,baker1989,faissner1978}
\begin{equation}
\frac{1}{E_{\nu}^{\text{kin}}}\overline{\sigma}(e^{-}\nu_{\mu}\rightarrow e^{-}\nu_{\mu})=(1.45\pm0.26)\times 10^{-42}~\text{cm}^{2}/\text{GeV}.
\end{equation}
\begin{table}[t]
\centering
\caption{\label{tab0}Numerical values of the laser-free DCS (Eq.~(\ref{freedcs})) taking into account the mass of muon neutrino for different neutrino kinetic energies. The initial and final angles are $\theta_{i}=1^{\circ}$ and $\theta_{f}=0^{\circ}$.}
\begin{ruledtabular}
\begin{tabular}{ccc}
&\multicolumn{2}{c}{Laser-free DCS~($10^{-36}~\text{eV}^{-2}$)} \\
\cline{2-3}
$E_{\nu}^{\text{kin}}$ (eV)& $m_{\nu_{\mu}}=0.15~\text{MeV}$ &$m_{\nu_{\mu}}=0$ \\
\hline
$0.1 $&$ 4.95  $ & $ 4.58$  \\
$0.5 $&$ 4.95  $ & $ 4.58$  \\
$1 $&$ 4.95 $ & $4.58$    \\
$5 $&$ 4.94 $ & $4.58$    \\
$10 $&$4.93 $ & $4.58$ \\
$50 $&$4.91 $ & $4.58$ \\
$10^{2} $&$4.89$ & $4.57$\\
$5\times10^{2} $&$4.83$ & $4.53$\\
$10^{3} $& $4.78$ & $4.47$   \\
$5\times10^{4} $ & $4.31$ & $4.15$   \\
$10^{5} $ & $4.24$ & $4.15$  \\
$5\times10^{5} $ & $4.17$ & $4.16$  \\
$10^{6} $ & $4.16$ & $4.16$ \\
$5\times10^{6} $ & $4.16$ & $4.16$ \\
$10^{7} $ & $4.16$ & $4.16$  \\
$5\times10^{7} $ & $4.16$ & $4.16$  \\
$10^{8}$ & $4.16$ & $4.16$  \\
$5\times10^{8}$ & $4.16$ & $4.16$  \\
$10^{9}$ & $4.16$ & $4.16$  \\
$10^{10}$ & $4.16$ & $4.16$  \\
$10^{14}$ & $2.85\times10^{12} $ & $2.85\times10^{12}$  \\
$10^{15}$ & $1.65\times10^{18} $ & $1.65\times10^{18}$  
\end{tabular}
\end{ruledtabular}
\end{table}
Instead of neglecting the neutrinos mass, let us now see what happens to the DCS if we consider their mass. Some of the changes that will occur in the theoretical calculation were summarized previously. In what follows, we will discuss the results of the nonzero neutrinos mass effect on the DCS in the absence of the laser and discover under what conditions this effect disappears. In Table \ref{tab0}, we list the values of the laser-free DCS as a function of the kinetic energy of the neutrino. The results for massless neutrinos are also included for comparison. The kinetic energy of the neutrino has been intentionally varied from its very low values to very high ones in order to make the effect of a nonzero neutrino mass more noticeable. From this table, one can clearly see the slight difference that exists between the DCS values at lower neutrino kinetic energies. The DCS when considering neutrino mass develops a small correction to its equivalent when neglecting neutrino mass, and thus it always remains the largest. With the increase of the neutrino kinetic energy, from value $E_{\nu}^{\text{kin}}=10^{6}~\text{eV}$, we see that the difference between the values disappears gradually until it completely vanishes at very high kinetic energies. This result could be used as a justification for choosing the value of the neutrino kinetic energy in the case of zero neutrino mass to $10^{6}~\text{eV}$ order of magnitude and beyond. Hereafter, we take the kinetic energy of the neutrino exactly as $E_{\nu}^{\text{kin}}=0.5\times10^{6}~\text{eV}$. 
\subsection{With laser field}
After a thorough discussion of the results obtained in the absence of a laser field, let us now turn to see what happens when we embed our scattering process (\ref{process}) in a circularly polarized electromagnetic field. The latter can be supplied experimentally by applying a laser instrument \cite{lasercp}. The direction of the field wave vector $\textbf{k}$ is along the $z$-axis, whereas the polarization vectors $\boldsymbol{\eta_1}$ and $\boldsymbol{\eta_2}$ perpendicular to $\textbf{k}$ are along the $x$- and $y$-axes, respectively. In the absence of a laser, our DCS depends on the total energies of the incoming electron and neutrino, as well as on the spherical angles. In the presence of the laser, three other variables are added to these ones; two of them, the field strength $\mathcal{E}_{0}$ and frequency $\omega$, characterize the laser and the third is the number of photons exchanged $n$ between the laser and the scattering system, which emerged due to the introduction of ordinary Bessel functions in the theoretical calculation. Therefore, to cover the laser-free results, it is sufficient, if the calculation is accurate, to cancel the number of photons $n$ and the field strength $\mathcal{E}_{0}$. Similarly, the effect of the laser on the DCS will be highlighted by studying its variations as a function of those parameters related to the electromagnetic field. Let us start by presenting the results related to how the laser affects the photons exchange phenomenon. 
\begin{figure}[hptb]
 \centering
\includegraphics[scale=0.53]{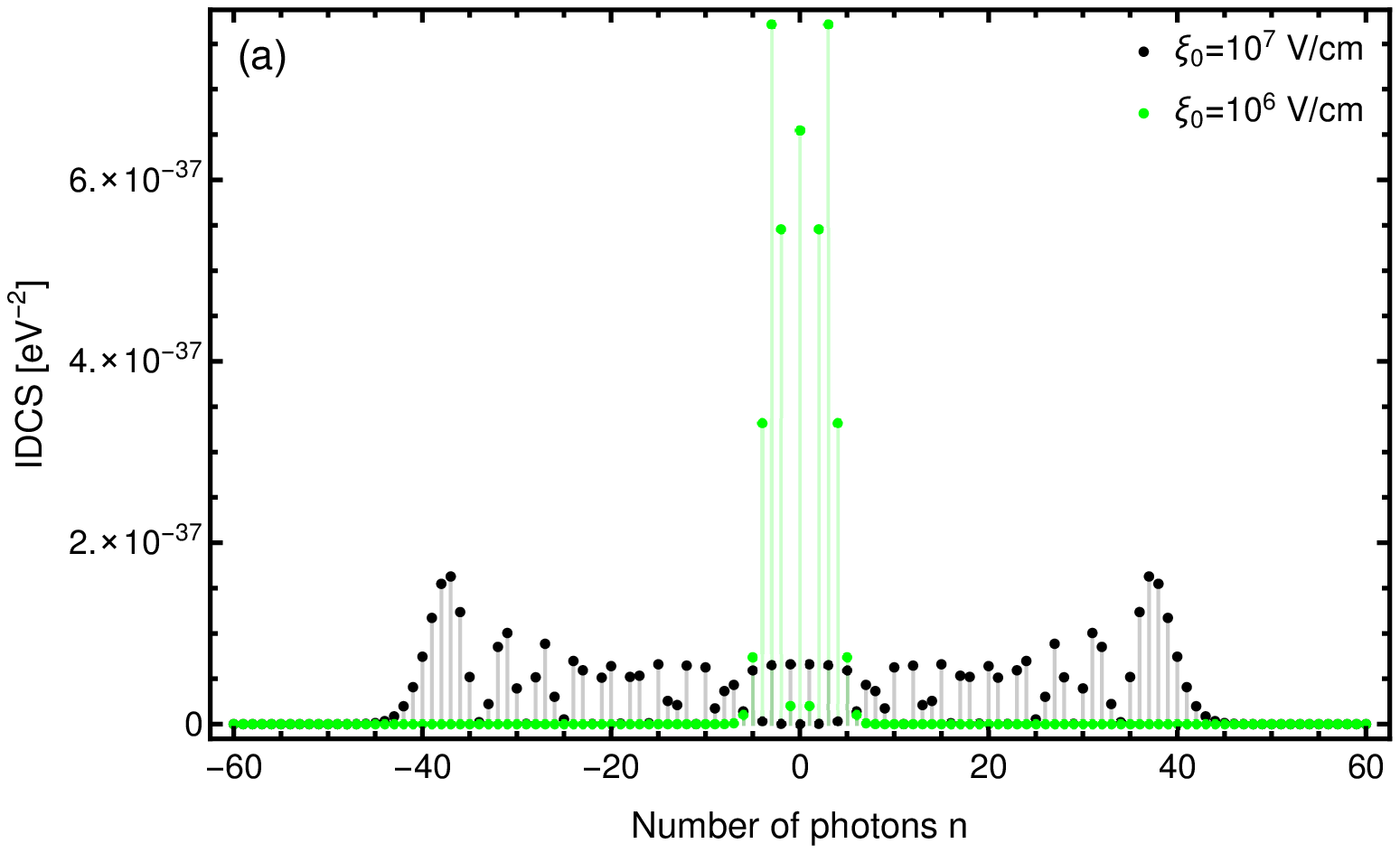}\hspace*{0.4cm}
\includegraphics[scale=0.54]{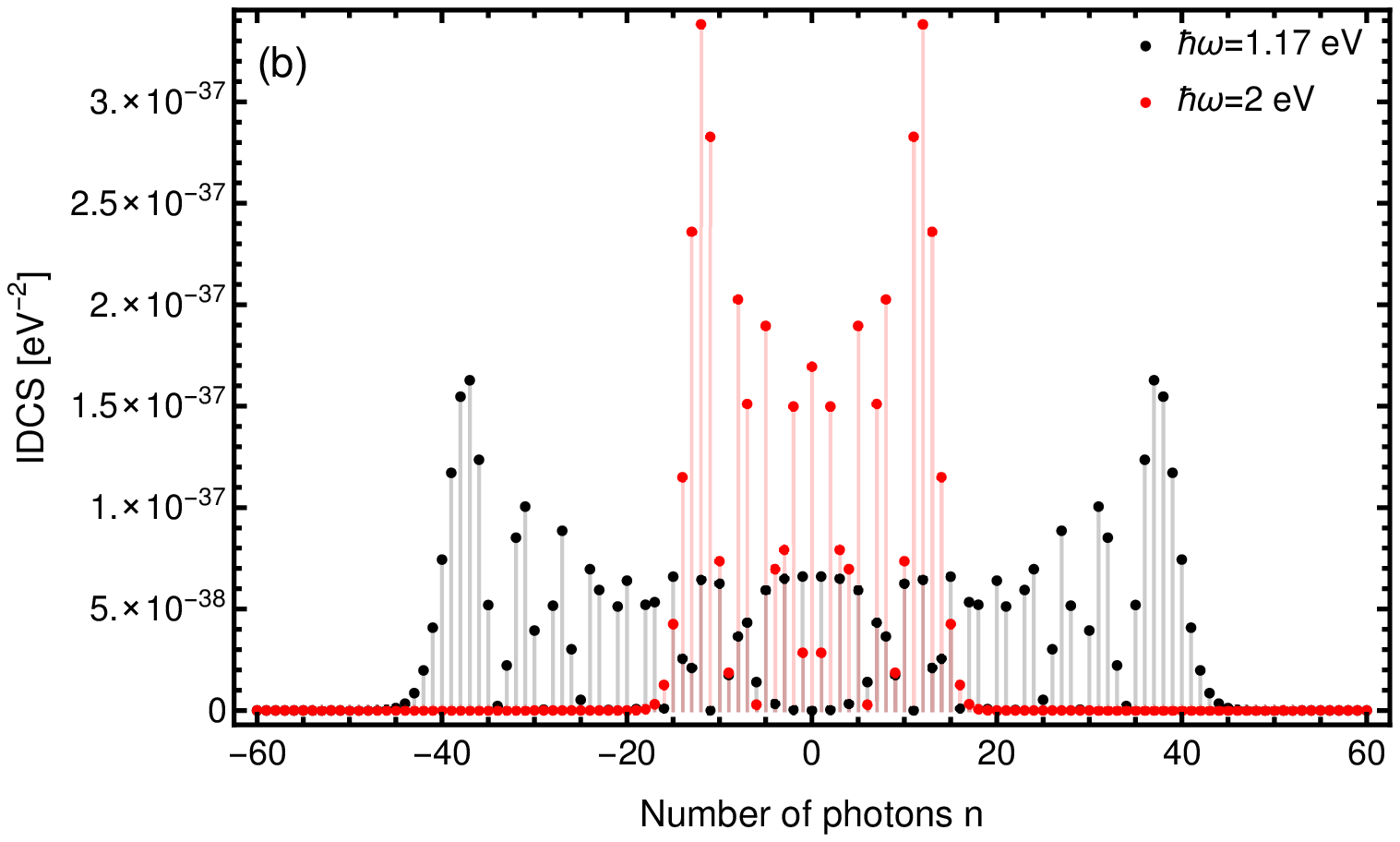}\par\vspace*{0.4cm}
\includegraphics[scale=0.53]{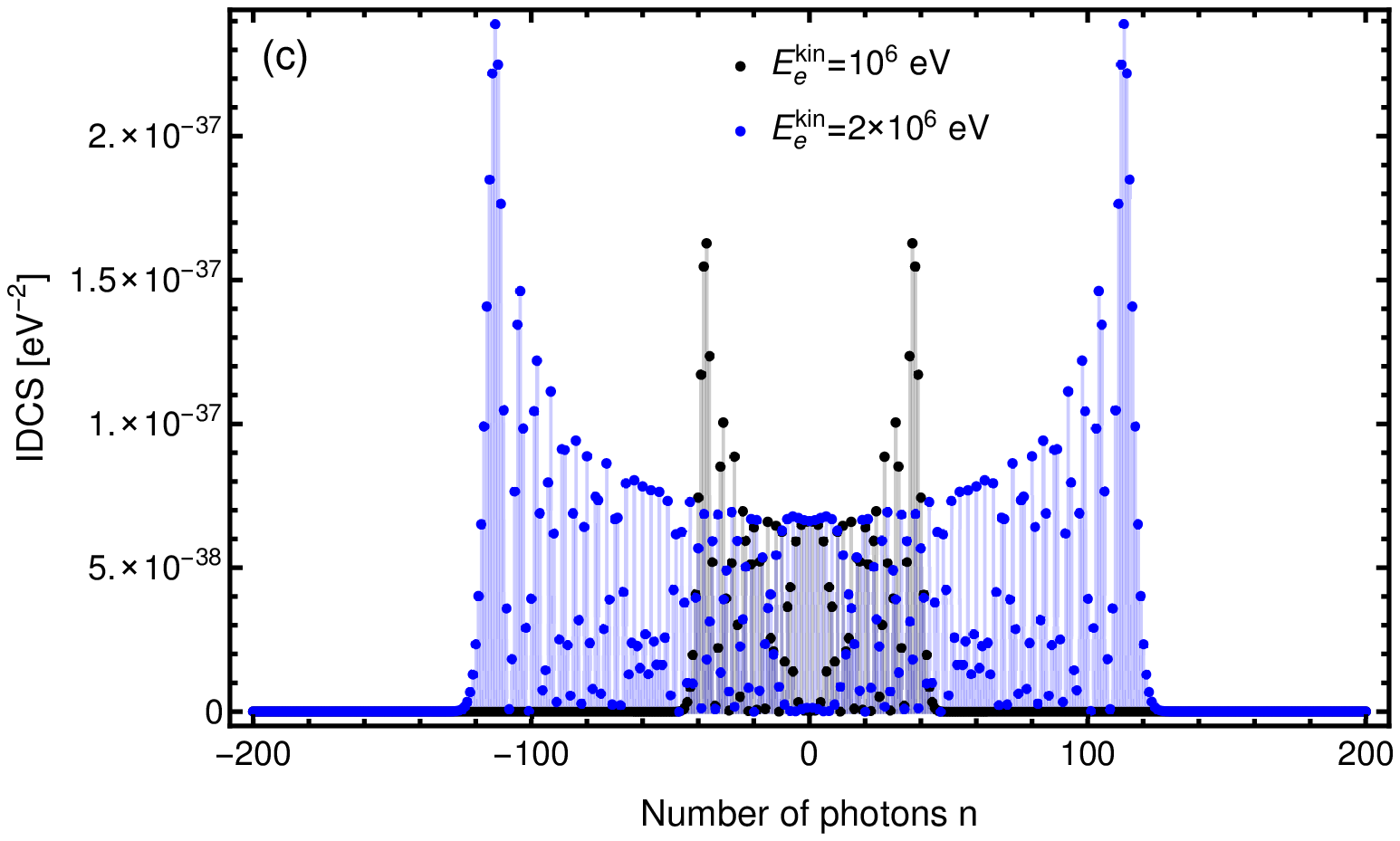}\hspace*{0.4cm}
\includegraphics[scale=0.53]{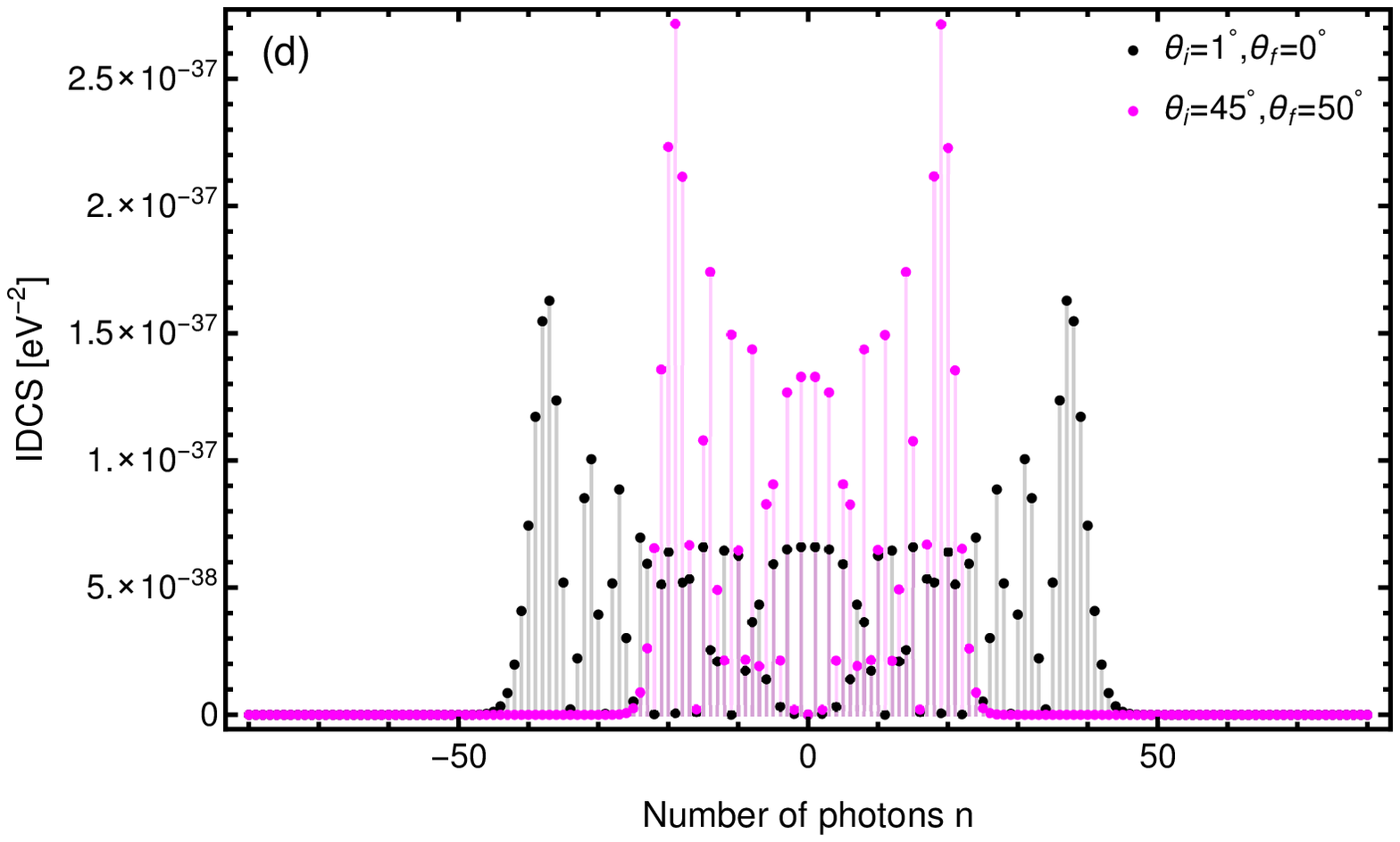}
\caption{The behavior of the IDCS, $d\overline{\sigma}^{n}/d\Omega$, versus the number of photons $n$. The different parameters are (a) $\hbar\omega=1.17$ eV, $\theta_{i}=1^{\circ}$ and $\theta_{f}=0^{\circ}$, (b) $\mathcal{E}_{0}=10^{7}$ V/cm, $\theta_{i}=1^{\circ}$ and $\theta_{f}=0^{\circ}$, (c) $\mathcal{E}_{0}=10^{7}$ V/cm, $\hbar\omega=1.17$ eV, $\theta_{i}=1^{\circ}$ and $\theta_{f}=0^{\circ}$ and (d) $\mathcal{E}_{0}=10^{7}$ V/cm and $\hbar\omega=1.17$ eV.}\label{envelopes}
\end{figure}
Figure~\ref{envelopes} depicts the effect of field strength $\mathcal{E}_{0}$ (Fig.~\ref{envelopes}\textcolor{blue}{(a)}), frequency $\hbar\omega$ (Fig.~\ref{envelopes}\textcolor{blue}{(b)}), electron kinetic energy (Fig.~\ref{envelopes}\textcolor{blue}{(c)}) and geometry (Fig.~\ref{envelopes}\textcolor{blue}{(d)}) on the photon exchange process. These graphs, which represent the changes in the IDCS in terms of photon number $n$, exhibit symmetrical envelopes with respect to the $n=0$-axis, which indicates that the photon emission processes $(n>0)$ are exactly equal to the photon absorption processes $(n<0)$. We did not join the points together because the number of photons $n$ is a discrete relative integer. The oscillations of these envelopes and their abrupt drop on the sides of each figure, as well as the positions of the peaks, are due to the well-known properties of Bessel functions \cite{szymanowski}. In Fig.~\ref{envelopes}\textcolor{blue}{(a)}, we show the IDCS changes in terms of the number of photons $n$ for two different field strengths, $\mathcal{E}_{0}=10^{6}$ V/cm and $\mathcal{E}_{0}=10^{7}$ V/cm, at a specific laser frequency $\hbar\omega=1.17$ eV. It seems clear to us that the photon exchange enhances with the increase of the field strength. In the case of $\mathcal{E}_{0}=10^{6}$ V/cm, only a few photons are exchanged and the cutoff number is about $n=\pm7$ compared to the case of $\mathcal{E}_{0}=10^{7}$ V/cm where the cutoff number is $n=\pm45$. A large number of exchanged photons indicates that the scattering process interacts strongly with the laser, which makes the effect of the latter on it important and prominent. In Fig.~\ref{envelopes}\textcolor{blue}{(b)}, the field strength is fixed at $10^{7}$ V/cm and the multiphoton processes are displayed for two different frequencies, $\hbar\omega=1.17$ eV and $\hbar\omega=2$ eV. It appears that at lower frequencies, the exchange of photons is important compared to higher frequencies. Based on the two Figs.~\ref{envelopes}\textcolor{blue}{(a)} and \ref{envelopes}\textcolor{blue}{(b)}, we conclude that the photon exchange process between the laser and the scattering process is related to the properties of the applied laser. Figure~\ref{envelopes}\textcolor{blue}{(c)} shows the dependence of the multiphoton phenomenon on the kinetic energy of the electron. We note that when the kinetic energy of the incoming electron increases, its interaction with the laser field enhances, which causes an exchange of a large number of photons between them. Figure~\ref{envelopes}\textcolor{blue}{(d)} illustrates how the chosen geometry can also affect the multiphoton processes. The number of exchanged photons in geometry $\theta_{i}=1^{\circ}$ and $\theta_{f}=0^{\circ}$ is greater than that exchanged in geometry $\theta_{i}=45^{\circ}$ and $\theta_{f}=50^{\circ}$. The difference between these two geometries is that the first means, according to well-known spherical coordinates, that the electron is incoming almost in the same direction as the field vector $\textbf{k}$, which is the $z$-axis direction, while in the second geometry the electron comes at an angle of $\theta_{i}=45^{\circ}$ to the $z$-axis. This indicates that the electron interacts with the laser if they are incoming together in the same direction more than if they are in two different directions.
\begin{figure}[hptb]
 \centering
\includegraphics[scale=0.6]{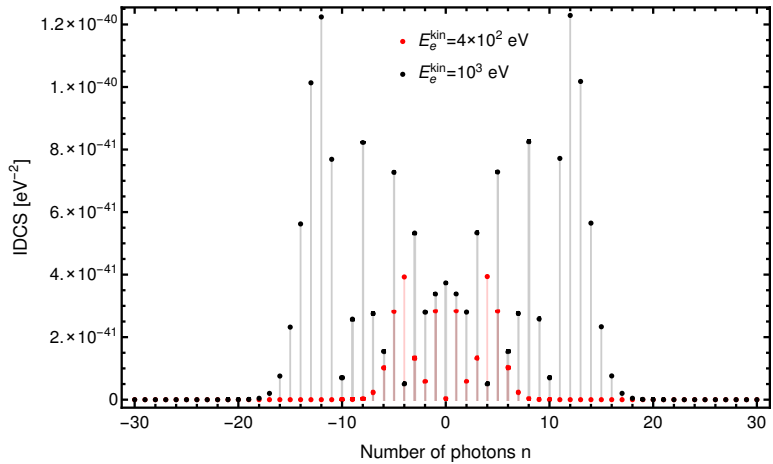}
\caption{The variations of the IDCS, $d\overline{\sigma}^{n}/d\Omega$, versus the number of photons $n$ for very low electron kinetic energies $E_e^{\text{kin}}=4\times 10^2$ and $10^3$ eV. The laser parameters are $\hbar\omega=1.17$ eV and $\mathcal{E}_{0}=10^{9}$ V/cm. The geometry parameters are $\theta_{i}=1^{\circ}$ and $\theta_{f}=0^{\circ}$.}\label{env1}
\end{figure}\\
In Fig.~\ref{env1}, we have included the result obtained in the case of very low electron kinetic energy in order to consider the situation where the electron is (at least approximately) at rest. This case is more natural and closer to realistic experimental conditions. From Fig.~\ref{env1}, it appears that in the case where the electron is almost at rest, it interacts little with the laser field, since it has exchanged very few photons, even at high field strengths ($10^{9}$ V/cm), compared to the case in which it has more kinetic energy. Therefore, the laser will not have a significant effect on an electron that is almost at rest. The effect of the laser at moderate intensities will be clear and significant only when the electron moves with a certain kinetic energy that is not neglected. For example at $\mathcal{E}_{0}=10^{7}$ V/cm and for the kinetic energy $E_e^{\text{kin}}=10^3$ eV, we found that the electron has emitted and absorbed only one photon ($n=\pm 1$).\\
\begin{figure}[hptb]
 \centering
\includegraphics[scale=0.5]{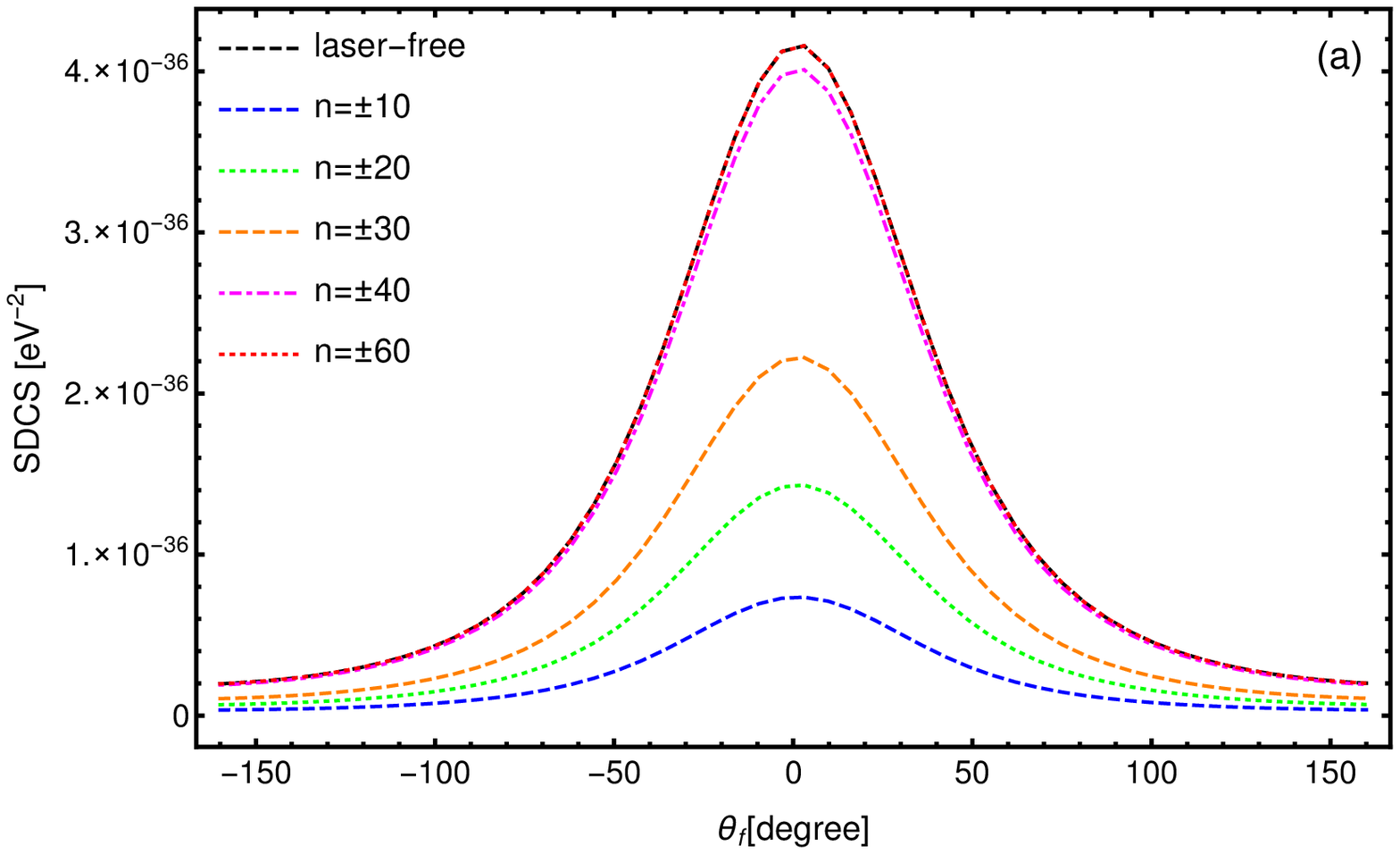}\hspace*{0.4cm}
\includegraphics[scale=0.5]{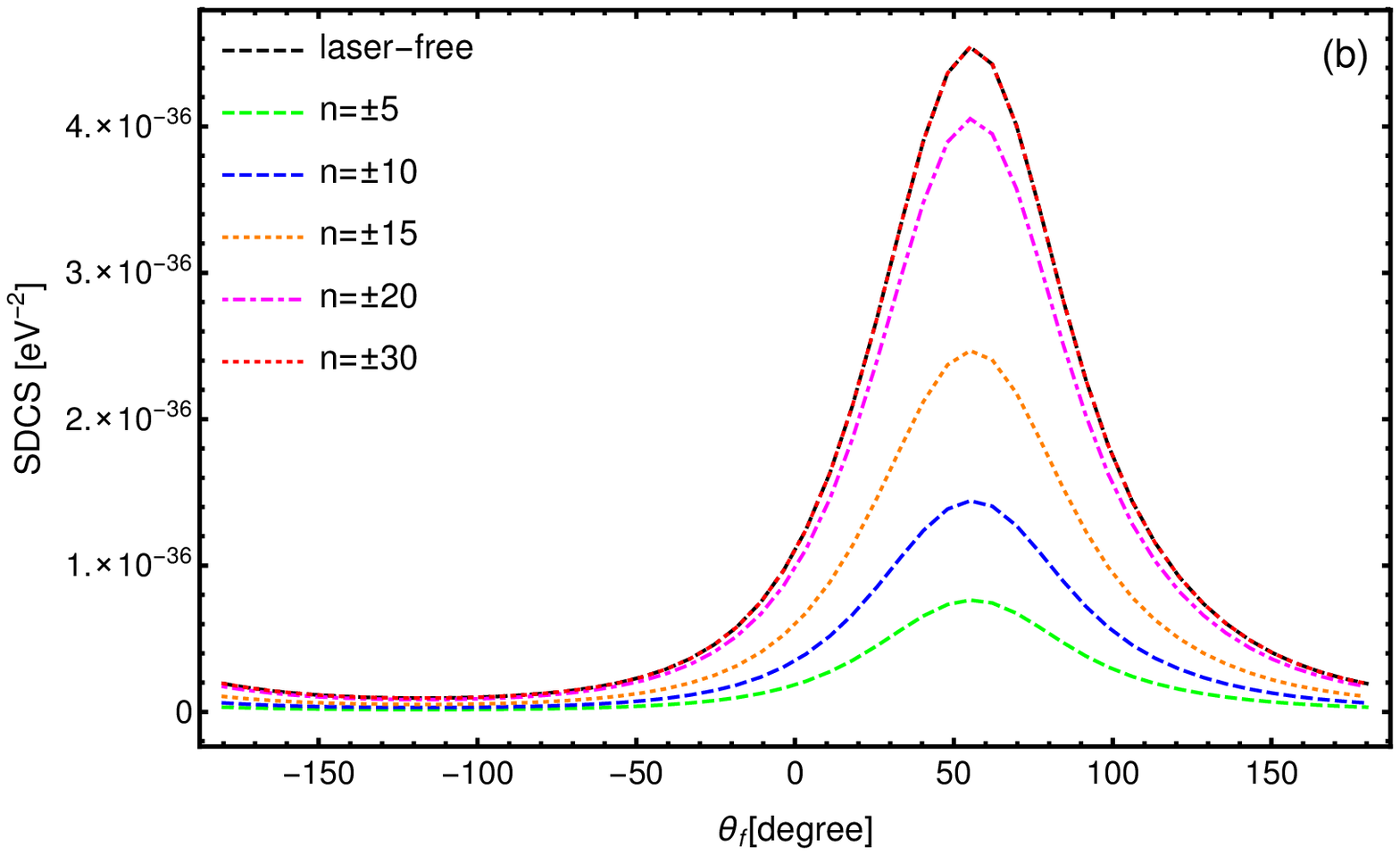}
\caption{The variations of the SDCS as a function of the final angle $\theta_{f}$ for various numbers of photons exchanged. The laser field strength and frequency are, respectively, $\mathcal{E}_{0}=10^{7}$ V/cm and $\hbar\omega=1.17$ eV. The initial angle is (a) $\theta_{i}=1^{\circ}$ and (b) $\theta_{i}=45^{\circ}$. The
notation $n=\pm N$ means that we have summed over the range of values $-N\leq n\leq+N$.}\label{sumrule}
\end{figure}
\begin{table}[h]
\centering
\caption{\label{tab1}Numerical values of the SDCS summed over a fixed number of photons $-20\leq n\leq+20$ in terms of field strength for three different frequencies. The initial and final angles are $\theta_{i}=1^{\circ}$ and $\theta_{f}=0^{\circ}$, respectively.}
\begin{ruledtabular}
\begin{tabular}{cccc}
&\multicolumn{3}{c}{SDCS~($10^{-36}~\text{eV}^{-2}$)} \\
\cline{2-4}
$\mathcal{E}_{0}~(\text{V/cm})$& $\hbar\omega= 0.117~\text{eV}$ &$\hbar \omega=1.17~\text{eV}$  &$\hbar\omega=2~\text{eV}$  \\
\hline
$10 $    &$ 4.16  $ & $ 4.16$ & $4.16  $ \\
$10^{2} $&$ 4.16 $ & $4.16 $ & $4.16 $   \\
$10^{3} $&$4.16$ & $4.16$ & $4.16$\\
$10^{4} $&$4.16 $ & $4.16  $ & $4.16$\\
$10^{5} $& $1.43 $ & $4.16 $ & $ 4.16$  \\
$10^{6} $ & $1.32\times10^{-1} $ & $ 4.16$ & $4.16$  \\
$10^{7} $ & $1.33\times10^{-2}  $ & $1.43$ & $ 4.16$   \\
$10^{8} $ & $1.35\times10^{-3} $ & $1.32\times10^{-1}  $ & $4.05\times10^{-1}$   \\
$10^{9} $ & $1.49\times10^{-4} $ & $1.37\times10^{-2}$ & $3.94\times10^{-2} $  \\
$10^{10}$ & $2.29\times10^{-5} $ & $1.50\times10^{-3}$ & $4.06\times10^{-3} $  \\
$10^{11}$ & $3.54\times10^{-6} $ & $2.36\times10^{-4}$ & $ 1.54\times10^{-3}$ 
\end{tabular}
\end{ruledtabular}
\end{table}
So far, we have discussed the variations of the IDCS in terms of the number of photons exchanged and reported the effect of different parameters on the photon exchange process. Let us now show the results of the laser-assisted DCS summed over a given number of photons. Figure~\ref{sumrule} presents the changes of the SDCS summed over a different numbers of exchanged photons as a function of the final scattering angle $\theta_{f}$. The variations of the laser-free DCS are also included for comparison. From this figure, we see that the laser-free DCS curve is always the highest. It is also noted that with the increase of the number of photons over which we summed, the SDCS tends towards the laser-free one so that they overlap perfectly when reaching the cutoff number. This can be translated mathematically as follows:
\begin{equation}\label{sumrule0}
\sum_{n=-\text{cutoff}}^{+\text{cutoff}}\dfrac{d\overline{\sigma}^{n}}{d\Omega}=\bigg(\dfrac{d\overline{\sigma}}{d\Omega}\bigg)^{laser-free}.
\end{equation}
The cutoff number differs from one geometry to another, being $n=\pm60$ in geometry $\theta_{i}=1^{\circ}$ and $n=\pm30$ in geometry $\theta_{i}=45^{\circ}$ (see Fig.~\ref{envelopes}\textcolor{blue}{(d)}). This convergence reached in these two figures is known by the so-called sum rule (Eq.~(\ref{sumrule0})) \cite{sumrule}, which has been previously confirmed to be fulfilled in several processes \cite{mouslih,mouslih1,imane,hrour,taj}.
At first, it was demonstrated in the atomic processes that occur in the presence of the laser in the nonrelativistic regime \cite{geltman,geltman1,kaminski,holmes}, and then it was applied to other relativistic scattering and weak decay processes. This sum rule is fulfilled only when summing over a specified number of photons, called the cutoff number, and above $(|n|\geq |\text{cutoff}|)$. The cutoff number is defined as a fixed number of photons from which the IDCS $d\overline{\sigma}^{n}/d\Omega$ suddenly falls off $(d\overline{\sigma}^{n}/d\Omega=0)$. It can be explained by the well-known properties of the Bessel function, which decreases considerably when its order $n$ is approximately equal to its argument $z$ [see Fig.~(\ref{fig9})]. Our main objective in drawing the envelopes shown in Fig.~(\ref{envelopes}) was to determine the cutoff number, at each specific field strength and frequency. In the context of laser-matter interaction, the only special functions included in the theoretical calculation are the Bessel functions. Consequently, their mathematical properties inevitably play an important role in the behavior of the calculated quantities, for example the SDCS here. The sum rule is thus, like the cutoff number, a reflection of the properties of Bessel functions. Let us give an example to illustrate this. If we look at the SDCS expression in Eq.~(\ref{dcswithlaser}) and the expression for the two traces given in the Appendix, we find that each term in the sum corresponds to a process in which a net number $n$ of photons are emitted (or absorbed for negative $n$). Let us take, for example, the first three terms of Eq.~(\ref{mfiappendix}) in the Appendix, which are multiplied by the square of the ordinary Bessel functions $J_n^2(z)$, $J_{n+1}^2(z)$ and $J_{n-1}^2(z)$. In Fig.~(\ref{fig9}), we give the changes of these functions in terms of order $n$ and for different arguments $z$. It is clear that the shape of the envelopes in this figure is the same as the one shown in Fig.~(\ref{envelopes}). It is these functions that give this shape to the envelopes that present the variations of the IDCS with respect to the number of photons exchanged $n$. The sum of these functions over order $n$ leads to two different cases. The first is the summation over an order $n$ less than the cutoff number $(|n|<|\text{cutoff}|)$ for each Bessel function, which gives
\begin{equation}
\sum_{-80}^{80}J_n^2(100)=0.588359;~~~\sum_{-100}^{100}J_{n+1}^2(150)=0.467531;~~~\sum_{-150}^{150}J_{n-1}^2(200)=0.541836.
\end{equation}
The second is the summation over an order $n$ greater than or equal to the cutoff number
$(|n|\geq |\text{cutoff}|)$
\begin{equation}
\sum_{-120}^{120}J_n^2(100)=1;~~~\sum_{-180}^{180}J_{n+1}^2(150)=1;~~~\sum_{-220}^{220}J_{n-1}^2(200)=1.
\end{equation}
If we focus as an example on $J_n^2(100)$, according to Fig.~(\ref{fig9}) its cutoff number is approximately equal to $100$ (or $-100$). The relationship $\sum_{-120}^{120}J_n^2(100)=1$ is in fact a kind of sum rule that brings us, together with the other functions, to the last one that we have included in Eq.~(\ref{sumrule0}); this means that summing the square of function $J_n(z)$, $J_n^2(z)$, over its cutoff number at a specific argument $z$ is exactly equal to its square when the order $n$ and the argument $z$ are zero (this indicates in our case the absence of a laser),
\begin{equation}
\sum_{-\text{cutoff}}^{+\text{cutoff}}J_n^2(z)=J_0^2(0)=1.
\end{equation}
One can connect this last equation with the one given in Eq.~(\ref{sumrule0}). All this is just to show how much the Bessel functions play a significant role in the results obtained.
\begin{figure}[hbtp]
\centering
\includegraphics[scale=0.75]{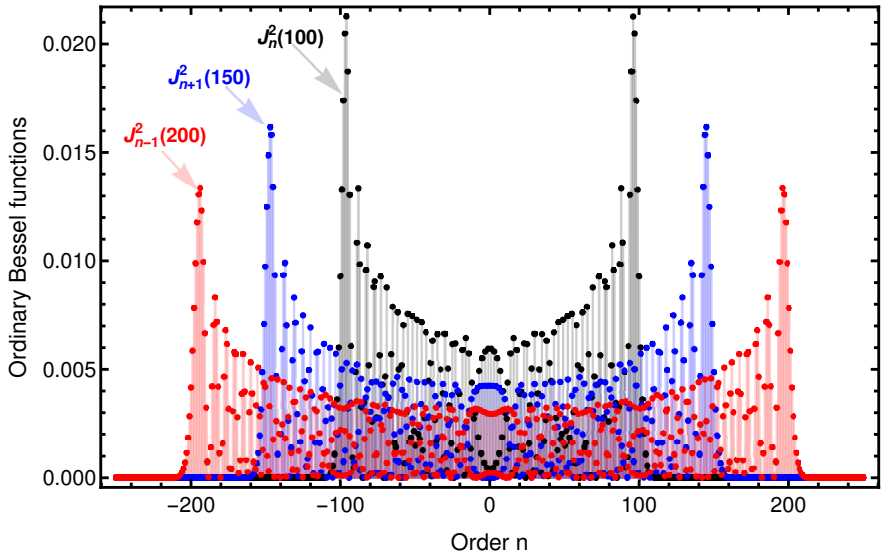}
\caption{The Bessel functions $J_n^2(z)$, $J_{n+1}^2(z)$ and $J_{n-1}^2(z)$  as a function of the order $n$, with different arguments $z$. The cutoff at $n\approx z$ is clearly visible.}\label{fig9}
\end{figure}\\
Table \ref{tab1} contains numerical values of the SDCS summed over a fixed number of photons $-20\leq n\leq+20$ in terms of field strength and for three different frequencies. It is evident from this table that the SDCS is not affected by the laser at low field strengths as it takes a constant value, but as soon as the laser becomes strong it begins to decrease. As for the frequency dependence, it is found that the effect of the laser diminishes with increasing frequency and that the low-frequency laser affects the SDCS faster than that of high frequency. As if the electron, at high frequencies, is not able to follow the electric field oscillations, and therefore the influence of the laser field on it decreases. The coupling of the electron to the laser field, and thus the effect of the laser, is theoretically determined by the argument of the ordinary Bessel functions, $z\propto \mathcal{E}_{0}/\omega^{2}$, defined in Eq.~(\ref{argument}). Thus, one would expect an increased effect of the laser on the SDCS, which implies an enhanced exchange of photons, for stronger fields or lower frequencies (see Figs.~\ref{envelopes}\textcolor{blue}{(a)} and \ref{envelopes}\textcolor{blue}{(b)}).
\begin{table}
\centering
\caption{\label{tab2}Numerical values of the SDCS summed over a fixed number of photons $-20\leq n\leq+20$ taking into account the mass of muon neutrino for different field strengths. The laser frequency is $\hbar\omega=1.17$ eV. The initial and final angles are $\theta_{i}=1^{\circ}$ and $\theta_{f}=0^{\circ}$. The neutrino kinetic energy is $E^{\text{kin}}_{\nu}=0.5\times10^{6}\text{eV}$.}
\begin{ruledtabular}
\begin{tabular}{ccc}
&\multicolumn{2}{c}{SDCS~($10^{-36}~\text{eV}^{-2}$)} \\
\cline{2-3}
$\mathcal{E}_{0}~(\text{V/cm})$& $m_{\nu_{\mu}}=0.15~\text{MeV}$ &$m_{\nu_{\mu}}=0$ \\
\hline
$10 $&$ 4.17  $ & $ 4.16 $  \\
$10^{2} $&$ 4.17$ & $ 4.16$  \\
$10^{3} $&$ 4.17$ & $4.16$    \\
$10^{4} $&$ 4.17$ & $4.16$    \\
$10^{5} $&$ 4.17$ & $4.16$ \\
$10^{6} $&$ 4.17$ & $4.16$ \\
$10^{7} $&$1.43$ & $1.43$\\
$10^{8} $&$1.32\times10^{-1}$ & $1.32\times10^{-1}$\\
$10^{9} $& $1.37\times10^{-2} $ & $1.37\times10^{-2}$ \\
$10^{10} $& $1.50\times10^{-3} $ & $1.50\times10^{-3} $ 
\end{tabular}
\end{ruledtabular}
\end{table}\\
The effect of the nonzero neutrino mass on the SDCS is displayed by the values listed in Table~\ref{tab2}, where we give numerical values for the SDCS, considering the mass of the muon neutrino, summed over a number of photons $-20\leq n\leq+20$ for different values of field strength. From this table, it is clear that the effect of the neutrino mass is noticeable only in the range of very weak field strengths. The same result was found by A. V. Kurilin when he studied the leptonic decays of the $W$ boson ($W^{\pm}\rightarrow \ell^{\pm}\bar{\nu}_{\ell}$) in the presence of a strong electromagnetic field \cite{kurilin2004}. He found, in the case of $m_{\nu}\neq0$, that the vacuum decay width correction develops a nontrivial oscillatory term, which can be served as an indication of the existence of massive neutrinos. Note that oscillations in the probabilities of quantum processes in an external field arise for a number of allowed reactions in vacuum if the participating particles have nonzero rest mass. For instance, it is found that there are oscillations that occur in the cross section for the process $\gamma e \rightarrow W \nu_{e}$ if the relevant calculations take into account the effects of a nonzero neutrino mass \cite{kurilin88}. In order to be more practical and accurate, we would like to point out here that the results presented in both Tables~\ref{tab1} and \ref{tab2} are more illustrative than realistic because fixing the number of photons and truncating the sum over it at a specific number, for example here $-20\leq n\leq+20$, does not give a realistic physical effect as long as we do not have any idea how many photons have actually been exchanged.
\section{Conclusion}\label{sec:conclusion}
Our goal in this paper was the theoretical investigation of the elastic scattering of a muon neutrino by an elctron in the presence of a circularly polarized monochromatic electromagnetic background. The DCS for this process has been analytically expressed in terms of the kinematical and laser parameters in both in the absence and presence of the laser. The results obtained show that the laser field has a considerable influence, via its strength and frequency, on the DCS as well as on the photon exchange process between the laser and the scattering system. Summing up the results, it can be concluded that the phenomenon of absorption and emission of multiple photons depends on the properties of the implemented laser as well as on the kinematical and geometrical conditions. Regarding the effect of the laser on the DCS, it was found that it increases with the field strength and gradually decreases at high frequencies. A sum rule, saying that the laser-assisted DCS summed over all the exchanged photon numbers gives the laser-free DCS, is found to be successfully fulfilled. The results, which took into account the neutrino mass, proved that the effect of a nonzero neutrino mass is noticeable only in the case of very low field strengths and neutrino kinetic energies. However, this approach based on Fermi theory, although adopted here and in a previous work \cite{bai2012} for the same process, may not be practical in all situations, especially at high energies, due to the well-known limitations of Fermi theory. Therefore, the next stage of our research will be a study of this process within the framework of electroweak theory, in which the relevant scattering process is mediated only by the exchange of the neutral $Z$ boson.
\appendix*
\section{Explicit expression of $\sum_{t_{i,f},s_{i,f}}|M^{n}_{fi}|^{2}$}\label{appendix}
The evaluation of the two traces shown in Eq.~(\ref{dcswithlaser}) gives the following result:
\begin{equation}\label{mfiappendix}
\begin{split}
\sum_{t_{i,f},s_{i,f}}|M^{n}_{fi}|^{2}=&A J_{n}^{2}(z)+BJ_{n+1}^{2}(z)+C J_{n-1}^{2}(z)+D J_{n}(z) J_{n+1}(z)+E J_{n}(z) J_{n-1}(z)\\&+F J_{n-1}(z) J_{n+1}(z),
\end{split}
\end{equation}
where the coefficients $A$, $B$, $C$, $D$, $E$ and $F$ are explicitly expressed, in terms of different 4-vector products, by:
\begin{equation}\label{coeffA}
\begin{split}
A=&\dfrac{32}{(k.p_{f})(k.p_{i})}\Big[ |\textbf{a}|^4 e^4 (g_{A}^2 + g_{V}^2) (k.k_{f})(k.k_{i}) + 2 (k.p_{f})(k.p_{i}) (2 g_{A}~g_{V} (-(k_{f}.p_{i}) (k_{i}.p_{f}) + (k_{f}.p_{f})\\
&\times(k_{i}.p_{i})) + g_{V}^2 ((k_{f}.p_{i})(k_{i}.p_{f})+(k_{f}.p_{f})(k_{i}.p_{i})-(k_{i}.k_{f}) m^2) + 
    g_{A}^2 ((k_{f}.p_{i})(k_{i}.p_{f}) + (k_{f}.p_{f})\\
&\times(k_{i}.p_{i})+ (k_{i}.k_{f}) m^2))+|\textbf{a}|^2 e^2 \Big(2 g_{A}~g_{V} (-(k_{i}.p_{i}) (k.k_{f}) (k.p_{f}) + (k_{f}.p_{i}) (k.k_{i})(k.p_{f})+(k_{i}.p_{f})\\
&\times(k.k_{f})(k.p_{i})-(k_{f}.p_{f})(k.k_{i}) (k.p_{i}))+g_{A}^2((k_{i}.p_{i})(k.k_{f})(k.p_{f})+(k_{f}.p_{i})(k.k_{i})(k.p_{f})+(k_{i}.p_{f})\\
&\times(k.k_{f})(k.p_{i})+(k_{f}.p_{f})(k.k_{i})(k.p_{i})-2(k_{i}.k_{f})(k.p_{f})(k.p_{i})-2(k.k_{f})(k.k_{i})m^2 -2 (k.k_{f})\\
&\times(k.k_{i})(p_{i}.p_{f}))+g_{V}^2 ((k_{i}.p_{i})(k.k_{f})(k.p_{f})+(k_{f}.p_{i})(k.k_{i})(k.p_{f}) +(k_{i}.p_{f})(k.k_{f})(k.p_{i})+(k_{f}.p_{f})\\
&\times(k.k_{i})(k.p_{i})-2 (k_{i}.k_{f})(k.p_{f})(k.p_{i})+2(k.k_{f})(k.k_{i})m^2-2(k.k_{f})(k.k_{i})(p_{i}.p_{f}))\Big)\Big],
\end{split} 
\end{equation}
\begin{equation}
\begin{split}
B=&\dfrac{4 e^2~|\textbf{a}|^2}{(k.p_{f})^2 (k.p_{i})^2}\big[4 (k.p_{f}) (k.p_{i})~ \big(- g_{A}^2 (k_{i}.p_{f}) (k.k_{f}) (k.p_{f}) + 2~g_{A}~g_{V}~(k_{i}.p_{f}) (k.k_{f}) (k.p_{f})\\
& - g_{V}^2~ (k_{i}.p_{f})(k.k_{f})(k.p_{f}) +  g_{A}^2~ (k_{i}.p_{i}) (k.k_{f}) (k.p_{f}) -2~g_{A}~g_{V}~(k_{i}.p_{i}) (k.k_{f}) (k.p_{f})\\
& +  g_{V}^2~(k_{i}.p_{i}) (k.k_{f}) (k.p_{f}) - (\eta_{1}.k_{f}) (\eta_{1}.p_{i})~ g_{A}^2 ~(k.k_{i})(k.p_{f}) -(\eta_{2}.k_{f}) (\eta_{2}.p_{i})~ g_{A}^2~(k.k_{i}) (k.p_{f})\\
&-2 (\eta_{1}.k_{f}) (\eta_{1}.p_{i})~g_{A}~g_{V}~(k.k_{i}) (k.p_{f})- 2 (\eta_{2}.k_{f}) (\eta_{2}.p_{i})~g_{A}~g_{V}~ (k.k_{i}) (k.p_{f})-(\eta_{1}.k_{f}) (\eta_{1}.p_{i})~ g_{V}^2\\
&\times (k.k_{i}) (k.p_{f}) - (\eta_{2}.k_{f}) (\eta_{2}.p_{i}) g_{V}^2~(k.k_{i}) (k.p_{f}) - g_{A}^2~(k_{f}.p_{f}) (k.k_{i}) (k.p_{f}) - 2~ g_{A}~g_{V}\\
&\times (k_{f}.p_{f}) (k.k_{i}) (k.p_{f})-g_{V}^2 (k_{f}.p_{f}) (k.k_{i}) (k.p_{f}) + g_{A}^2 (k_{f}.p_{i})\times (k.k_{i}) (k.p_{f}) \\
&+ 2 g_{A} g_{V} (k_{f}.p_{i}) (k.k_{i}) (k.p_{f}) + g_{V}^2 (k_{f}.p_{i}) (k.k_{i}) (k.p_{f})+ g_{A}^2 (k_{i}.p_{f}) (k.k_{f}) (k.p_{i}) \\
&+ 2  g_{A} g_{V} (k_{i}.p_{f}) (k.k_{f}) (k.p_{i})+ g_{V}^2 (k_{i}.p_{f}) (k.k_{f}) (k.p_{i})- g_{A}^2 (k_{i}.p_{i}) (k.k_{f}) (k.p_{i}) \\
&- 2 g_{A} g_{V} (k_{i}.p_{i}) (k.k_{f}) (k.p_{i}) -  g_{V}^2 (k_{i}.p_{i}) (k.k_{f}) (k.p_{i}) + g_{A}^2 (k_{f}.p_{f}) (k.k_{i}) (k.p_{i}) \\
&- 2  g_{A} g_{V} (k_{f}.p_{f}) (k.k_{i})(k.p_{i})+ g_{V}^2 (k_{f}.p_{f}) (k.k_{i}) (k.p_{i})- g_{A}^2 (k_{f}.p_{i}) (k.k_{i}) (k.p_{i})\\
&+2  g_{A} g_{V} (k_{f}.p_{i}) (k.k_{i}) (k.p_{i}) - g_{V}^2(k_{f}.p_{i}) (k.k_{i}) (k.p_{i}) -2  g_{A}^2 (k_{i}.k_{f}) (k.p_{f}) (k.p_{i}) \\
&- 2  g_{V}^2 (k_{i}.k_{f}) (k.p_{f}) (k.p_{i}) +(\eta_{1}.p_{f}) (k.k_{i}) (2 (\eta_{1}.p_{i})(g_{A}^2 + g_{V}^2) (k.k_{f}) - (\eta_{1}.k_{f}) (g_{A} - g_{V})^2\\
&\times (k.p_{i}))+(\eta_{2}.p_{f}) (k.k_{i}) (2 (\eta_{2}.p_{i}) (g_{A}^2 + g_{V}^2) (k.k_{f}) - (\eta_{2}.k_{f})(g_{A} - g_{V})^2 (k.p_{i}))-2  g_{A}^2 \\
&\times(k.k_{f})(k.k_{i}) m^2 + 2 g_{V}^2 (k.k_{f}) (k.k_{i}) m^2 -2  (g_{A}^2 + g_{V}^2) (k.k_{f}) (k.k_{i}) (p_{i}.p_{f})\big)-2g_{A} g_{V}\\
&\times \big((k_{i}.p_{i}) (k.p_{f}) ((k.p_{f})^2 - 2 (k.p_{f}) (k.p_{i}) + 2 (k.p_{i})^2) - ((k.p_{f})-(k.p_{i})) ((k_{i}.p_{f}) (k.p_{f}) (k.p_{i})\\
& + (k.k_{i}) ((k.p_{f}) + (k.p_{i})) (p_{i}.p_{f}))\big)\epsilon(\eta_{1},\eta_{2},k,k_{f})-2 ~g_{A} g_{V} \big((k_{f}.p_{i}) (k.p_{f}) ((k.p_{f})^2 - 2 (k.p_{f}) \\
&\times(k.p_{i}) + 2 (k.p_{i})^2) - ((k.p_{f}) -(k.p_{i})) ((k_{f}.p_{f}) (k.p_{f}) (k.p_{i}) + (k.k_{f}) ((k.p_{f}) + (k.p_{i})) (p_{i}.p_{f})\big)\\
&\times\epsilon(\eta_{1},\eta_{2},k,k_{i})+2g_{A} g_{V}\epsilon(\eta_{1},\eta_{2},k,p_{f}) \big(2(k_{i}.k_{f}) (k.p_{f})^2 (k.p_{i})  - (k_{i}.p_{i}) (k.k_{f}) (k.p_{i})^2\\
&-(k_{f}.p_{i}) (k.k_{i}) (k.p_{i})^2 + (k_{i}.k_{f}) (k.p_{f}) (k.p_{i})^2\big) + 2 ~g_{A} g_{V}\epsilon(\eta_{1},\eta_{2},k,p_{i}) \big((k_{i}.p_{f}) (k.k_{f}) (k.p_{f})^2 \\
&+  (k_{f}.p_{f}) (k.k_{i}) (k.p_{f})^2  + 2  (k_{i}.k_{f}) (k.p_{f})^3  - 4  (k_{i}.k_{f}) (k.p_{f})^2 (k.p_{i}) \big) +\epsilon(\eta_{1},\eta_{2},k_{f},k_{i})\\
&\times\big(4 g_{A}^2 (k.p_{f})^3 (k.p_{i})+4 g_{V}^2 (k.p_{f})^3 (k.p_{i})+4g_{A}^2 (k.p_{f})  (k.p_{i})^3+4 g_{V}^2 (k.p_{f}) (k.p_{i})^3\big)- 2g_{A} g_{V} \\
&\times\epsilon(\eta_{1},\eta_{2},k_{f},p_{f})\big( (k.k_{i}) (k.p_{f})^2 (k.p_{i})+(k.k_{i}) (k.p_{f}) (k.p_{i})^2 + (k.k_{i}) (k.p_{i})^3 \big) +4 g_{A} g_{V} (k.k_{i})\\
&\times (k.p_{f})^2 (k.p_{i}) \epsilon(\eta_{1},\eta_{2},k_{f},p_{i})-2 ~g_{A} g_{V}\epsilon(\eta_{1},\eta_{2},k_{i},p_{f})\big((k.k_{f}) (k.p_{f})^2(k.p_{i}) + (k.k_{f}) \\
&\times(k.p_{f}) (k.p_{i})^2+(k.k_{f}) \big)+4 g_{A} g_{V}(k.k_{f}) (k.p_{f})^2 (k.p_{i}) \epsilon(\eta_{1},\eta_{2},k_{i},p_{i})-4~ \epsilon(\eta_{1},k,k_{f},k_{i})\\
&\times\big((\eta_{2}.p_{f}) g_{A}^2 (k.p_{f})^2 (k.p_{i}) + (\eta_{2}.p_{f}) g_{V}^2 (k.p_{f})^2 (k.p_{i})+ (\eta_{2}.p_{i}) g_{A}^2 (k.p_{f}) (k.p_{i})^2+ (\eta_{2}.p_{i})\\
&\times g_{V}^2 (k.p_{f}) (k.p_{i})^2\big)- g_{A} g_{V}~\epsilon(\eta_{1},k,k_{f},p_{f})\big((\eta_{2}.p_{i})(k.k_{i}) (k.p_{f})^2  - 4 (\eta_{2}.p_{i}) (k.k_{i}) (k.p_{f}) (k.p_{i})\\
&- (\eta_{2}.p_{i}) (k.k_{i}) (k.p_{i})^2\big) -g_{A} g_{V}~\epsilon(\eta_{1},k,k_{f},p_{i})\big((\eta_{2}.p_{f}) (k.k_{i}) (k.p_{f})^2 + 4 (\eta_{2}.p_{f}) (k.k_{i}) (k.p_{f})\\
&\times (k.p_{i})- (\eta_{2}.p_{f}) (k.k_{i}) (k.p_{i})^2\big)-g_{A} g_{V}~~ \epsilon(\eta_{1},k,k_{i},p_{f})\big((\eta_{2}.p_{i}) (k.k_{f}) (k.p_{f})^2- 4 (\eta_{2}.p_{i}) (k.k_{f})\\
&\times (k.p_{f}) (k.p_{i})- 2 (\eta_{2}.k_{f}) (k.p_{f})^2 (k.p_{i}) - (\eta_{2}.p_{i}) (k.k_{f}) (k.p_{i})^2-2 (\eta_{2}.k_{f}) (k.p_{f}) (k.p_{i})^2\big)\\
&- g_{A} g_{V}~\epsilon(\eta_{1},k,k_{i},p_{i})\big((\eta_{2}.p_{f}) (k.k_{f})(k.p_{f})^2 -2 (\eta_{2}.k_{f}) (k.p_{f})^3+ 4 (\eta_{2}.p_{f}) (k.k_{f}) (k.p_{f}) (k.p_{i})\\
&-(\eta_{2}.p_{f}) (k.k_{f}) (k.p_{i})^2\big)- g_{A}~g_{V}~~\epsilon(\eta_{1},k,p_{f},p_{i}) \big((\eta_{2}.k_{f}) (k.k_{i}) (k.p_{f})^2 + (\eta_{2}.k_{f}) (k.k_{i}) (k.p_{i})^2\big) \\
&+ \epsilon(\eta_{2},k,k_{f},k_{i})\big(4  g_{A}^2 (\eta_{1}.p_{f}) (k.p_{f})^2 (k.p_{i})+ 4 g_{V}^2 (\eta_{1}.p_{f}) (k.p_{f})^2(k.p_{i}) + 4  g_{A}^2 (\eta_{1}.p_{i}) (k.p_{f}) \\
&\times(k.p_{i})^2 + 4 g_{V}^2 (\eta_{1}.p_{i})  (k.p_{f}) (k.p_{i})^2\big)+  g_{A}~g_{V}~~ \epsilon(\eta_{2},k,k_{f},p_{f})\big((\eta_{1}.p_{i}) (k.k_{i}) (k.p_{f})^2\\
&-4 (\eta_{1}.p_{i}) (k.k_{i})(k.p_{f}) (k.p_{i})-(\eta_{1}.p_{i}) (k.k_{i}) (k.p_{i})^2\big)+ g_{A}~g_{V}~~\epsilon(\eta_{2},k,k_{f},p_{i})\big((\eta_{1}.p_{f})\\
&\times (k.k_{i}) (k.p_{f})^2+ 4 (\eta_{1}.p_{f}) (k.k_{i}) (k.p_{f}) (k.p_{i}) - (\eta_{1}.p_{f}) (k.k_{i}) (k.p_{i})^2\big)+ g_{A}~g_{V}~~\\
&\times\epsilon(\eta_{2},k,k_{i},p_{f})\big((\eta_{1}.p_{i})  (k.k_{f}) (k.p_{f})^2 -4 (\eta_{1}.p_{i}) (k.k_{f}) (k.p_{f}) (k.p_{i})-2 (\eta_{1}.k_{f}) (k.p_{f})^2 (k.p_{i})\\
&-(\eta_{1}.p_{i}) (k.k_{f}) (k.p_{i})^2+2 (\eta_{1}.k_{f}) (k.p_{f}) (k.p_{i})^2\big)+g_{A}~g_{V}~\epsilon(\eta_{2},k,k_{i},p_{i})\big((\eta_{1}.p_{f}) (k.k_{f}) \\
&\times(k.p_{f})^2 -2 (\eta_{1}.k_{f}) (k.p_{f})^3 +4 (\eta_{1}.p_{f}) (k.k_{f}) (k.p_{f})(k.p_{i})- (\eta_{1}.p_{f})(k.k_{f}) (k.p_{i})^2\big) \\
&+(\eta_{1}.k_{f}) g_{A} g_{V} (k.k_{i})((k.p_{f})^2+ (k.p_{i})^2)\epsilon(\eta_{2},k,p_{f},p_{i})\big],
\end{split} 
\end{equation}
\begin{equation}
\begin{split}
C=&\dfrac{4 e^2~|\textbf{a}|^2}{(k.p_{f})^2 (k.p_{i})^2}\big[4  (k.p_{f}) (k.p_{i})\big(- g_{A}^2 (k_{i}.p_{f}) (k.k_{f}) (k.p_{f}) + 2~g_{A}~g_{V} (k_{i}.p_{f}) (k.k_{f}) (k.p_{f})-g_{V}^2 (k_{i}.p_{f})\\
& \times(k.k_{f}) (k.p_{f}) + g_{A}^2 (k_{i}.p_{i}) (k.k_{f}) (k.p_{f}) -2 g_{A}~g_{V} (k_{i}.p_{i}) (k.k_{f}) (k.p_{f}) + g_{V}^2 (k_{i}.p_{i}) (k.k_{f}) (k.p_{f})\\
&-(\eta_{1}.k_{f}) (\eta_{1}.p_{i}) g_{A}^2 (k.k_{i}) (k.p_{f}) - (\eta_{2}.k_{f}) (\eta_{2}.p_{i}) g_{A}^2 (k.k_{i}) (k.p_{f})-2 (\eta_{1}.k_{f}) (\eta_{1}.p_{i}) g_{A} g_{V} (k.k_{i}) (k.p_{f}) \\
&- 2 (\eta_{2}.k_{f}) (\eta_{2}.p_{i}) g_{A} g_{V} (k.k_{i}) (k.p_{f})-(\eta_{1}.k_{f}) (\eta_{1}.p_{i}) g_{V}^2 (k.k_{i}) (k.p_{f}) - (\eta_{2}.k_{f}) (\eta_{2}.p_{i}) g_{V}^2 (k.k_{i}) (k.p_{f})\\
&- g_{A}^2 (k_{f}.p_{f}) (k.k_{i}) (k.p_{f}) - 2  g_{A} g_{V} (k_{f}.p_{f}) (k.k_{i}) (k.p_{f})- g_{V}^2  (k_{f}.p_{f}) (k.k_{i}) (k.p_{f}) + g_{A}^2 (k_{f}.p_{i}) (k.k_{i})\\
&\times (k.p_{f}) +2 g_{A} g_{V} (k_{f}.p_{i}) (k.k_{i}) (k.p_{f}) + g_{V}^2 (k_{f}.p_{i}) (k.k_{i}) (k.p_{f})+ g_{A}^2 (k_{i}.p_{f}) (k.k_{f}) (k.p_{i}) + 2 g_{A} g_{V}\\
&\times (k_{i}.p_{f}) (k.k_{f}) (k.p_{i})+ g_{V}^2 (k_{i}.p_{f}) (k.k_{f}) (k.p_{i})- g_{A}^2 (k_{i}.p_{i}) (k.k_{f}) (k.p_{i}) -2 g_{A} g_{V} (k_{i}.p_{i}) (k.k_{f})\\
& \times(k.p_{i})- g_{V}^2 (k_{i}.p_{i}) (k.k_{f}) (k.p_{i}) + g_{A}^2 (k_{f}.p_{f}) (k.k_{i}) (k.p_{i}) - 2 g_{A} g_{V} (k_{f}.p_{f}) (k.k_{i}) (k.p_{i})+ g_{V}^2 (k_{f}.p_{f})\\
&\times (k.k_{i}) (k.p_{i})-g_{A}^2 (k_{f}.p_{i}) (k.k_{i}) (k.p_{i})+2 g_{A} g_{V} (k_{f}.p_{i}) (k.k_{i}) (k.p_{i})- g_{V}^2 (k_{f}.p_{i}) (k.k_{i}) (k.p_{i})\\
&-2g_{A}^2 (k_{i}.k_{f}) (k.p_{f}) (k.p_{i}) - 2  g_{V}^2 (k_{i}.k_{f}) (k.p_{f}) (k.p_{i})+ (\eta_{1}.p_{f}) (k.k_{i}) (2 (\eta_{1}.p_{i}) (g_{A}^2 + g_{V}^2) (k.k_{f})\\
& - (\eta_{1}.k_{f}) (g_{A} - g_{V})^2 (k.p_{i}))+(\eta_{2}.p_{f}) (k.k_{i}) (2 (\eta_{2}.p_{i}) (g_{A}^2 + g_{V}^2) (k.k_{f}) - (\eta_{2}.k_{f}) (g_{A} - g_{V})^2 (k.p_{i}))\\
&-2 g_{A}^2 (k.k_{f}) (k.k_{i}) m^2 + 2 g_{V}^2 (k.k_{f}) (k.k_{i}) m^2 -2 (g_{A}^2 + g_{V}^2) (k.k_{f}) (k.k_{i})  (p_{i}.p_{f}) \big) +2~g_{A}~g_{V} \big((k_{i}.p_{i}) \\
&\times(k.p_{f}) ((k.p_{f})^2 - 2 (k.p_{f})(k.p_{i}) + 2 (k.p_{i})^2) - ((k.p_{f}) - (k.p_{i})) ((k_{i}.p_{f}) (k.p_{f}) (k.p_{i}) + (k.k_{i}) \\
&\times((k.p_{f}) + (k.p_{i})) (p_{i}.p_{f}))\big)\epsilon(\eta_{1},\eta_{2},k,k_{f})+2~g_{A}~g_{V}\big((k_{f}.p_{i}) (k.p_{f}) ((k.p_{f})^2 -2(k.p_{f}) (k.p_{i}) \\
&+ 2 (k.p_{i})^2)-((k.p_{f}) - (k.p_{i})) ((k_{f}.p_{f}) (k.p_{f}) (k.p_{i}) +(k.k_{f}) ((k.p_{f}) + (k.p_{i})) (p_{i}.p_{f}))\big)\\
&\times~\epsilon(\eta_{1},\eta_{2},k,k_{i})+2~g_{A}~g_{V}~\big(-2~(k_{i}.k_{f}) (k.p_{f})^2 (k.p_{i})+ (k_{i}.p_{i}) (k.k_{f}) (k.p_{i})^2+ (k_{f}.p_{i}) (k.k_{i})\\
& \times(k.p_{i})^2-2~ (k_{i}.k_{f}) (k.p_{f}) (k.p_{i})^2\big) \epsilon(\eta_{1},\eta_{2},k,p_{f}) +2~g_{A}~g_{V}\big(-(k_{i}.p_{f}) (k.k_{f}) ((k.p_{f})^2) \\
&-(k_{f}.p_{f}) (k.k_{i}) (k.p_{f})^2-2 (k_{i}.k_{f}) (k.p_{f})^3+4~(k_{i}.k_{f}) (k.p_{f})^2 (k.p_{i})\big)~\epsilon(\eta_{1},\eta_{2},k,p_{i}) \\
&+4~\big (-g_{A}^2 (k.p_{f})^3 (k.p_{i})- g_{V}^2 (k.p_{f})^3(k.p_{i})-g_{A}^2 (k.p_{f}) (k.p_{i})^3-g_{V}^2 (k.p_{f}) (k.p_{i})^3\big)\\
& \times\epsilon(\eta_{1},\eta_{2},k_{f},k_{i})+2~g_{A}~g_{V}~\big((k.k_{i}) (k.p_{f})^2 (k.p_{i})+(k.k_{i}) (k.p_{f}) (k.p_{i})^2+(k.k_{i}) (k.p_{i})^3\big) \\
&\times\epsilon(\eta_{1},\eta_{2},k_{f},p_{f}) -4~~g_{A}~g_{V} (k.k_{i}) (k.p_{f})^2 (k.p_{i})~\epsilon(\eta_{1},\eta_{2},k_{f},p_{i}) +2~g_{A}~g_{V}\big((k.k_{f}) (k.p_{f})^2 (k.p_{i})\\
&+ (k.k_{f}) (k.p_{f}) (k.p_{i})^2+(k.k_{f}) (k.p_{i})^3\big)\epsilon(\eta_{1},\eta_{2},k_{i},p_{f}) -4~~g_{A}~g_{V} (k.k_{f}) (k.p_{f})^2 (k.p_{i})~ \epsilon(\eta_{1},\eta_{2},k_{i},p_{i})\\
&+ 4~\big( g_{A}^2 (\eta_{2}.p_{f}) (k.p_{f})^2 (k.p_{i}) +   g_{V}^2 (\eta_{2}.p_{f}) (k.p_{f})^2 (k.p_{i})+ g_{A}^2 (\eta_{2}.p_{i}) (k.p_{f}) (k.p_{i})^2+  g_{V}^2 (\eta_{2}.p_{i}) (k.p_{f})\\
&\times (k.p_{i})^2\big)~\epsilon(\eta_{1},k,k_{f},k_{i})+ ~g_{A}~g_{V}~(\eta_{2}.p_{i})~\big((k.k_{i}) (k.p_{f})^2-4~(k.k_{i}) (k.p_{f}) (k.p_{i})-(k.k_{i}) (k.p_{i})^2 \big)\\
&\times\epsilon(\eta_{1},k,k_{f},p_{f})+ ~g_{A}~g_{V}~(\eta_{2}.p_{f})~\big((k.k_{i}) (k.p_{f})^2+4~(k.k_{i}) (k.p_{f}) (k.p_{i}) -(k.k_{i}) (k.p_{i})^2\big)\\
& \times\epsilon(\eta_{1},k,k_{f},p_{i})+ ~g_{A}~g_{V}\big((\eta_{2}.p_{i}) (k.k_{f}) (k.p_{f})^2-4 (\eta_{2}.p_{i}) (k.k_{f}) (k.p_{f}) (k.p_{i})-2 (\eta_{2}.k_{f})(k.p_{f})^2 (k.p_{i})\\
&-(\eta_{2}.p_{i}) (k.k_{f}) (k.p_{i})^2+2 (\eta_{2}.k_{f}) (k.p_{f}) (k.p_{i})^2\big)~\epsilon(\eta_{1},k,k_{i},p_{f})+~g_{A}~g_{V}~\big((\eta_{2}.p_{f}) (k.k_{f}) (k.p_{f})^2\\
&-2 (\eta_{2}.k_{f}) (k.p_{f})^3+4 (\eta_{2}.p_{f}) (k.k_{f}) (k.p_{f}) (k.p_{i})-(\eta_{2}.p_{f}) (k.k_{f}) (k.p_{i})^2\big)~\epsilon(\eta_{1},k,k_{i},p_{i})\\
&+ ~g_{A}~g_{V}~(\eta_{2}.k_{f})\big( (k.k_{i}) (k.p_{f})^2+(k.k_{i}) (k.p_{i})^2\big)~\epsilon(\eta_{1},k,p_{f},p_{i}) + 4\big(-~g_{A}^2~(\eta_{1}.p_{f}) (k.p_{f})^2 (k.p_{i})\\
&-g_{V}^2~ (\eta_{1}.p_{f}) (k.p_{f})^2 (k.p_{i})- g_{A}^2 (\eta_{1}.p_{i})  (k.p_{f}) (k.p_{i})^2 -g_{V}^2~(\eta_{1}.p_{i}) (k.p_{f}) (k.p_{i})^2\big)~\epsilon(\eta_{2},k,k_{f},k_{i})\\
&+ ~g_{A}~g_{V}~(\eta_{1}.p_{i})\big(-(k.k_{i}) (k.p_{f})^2+4~(k.k_{i}) (k.p_{f}) (k.p_{i})+(k.k_{i}) (k.p_{i})^2\big)~\epsilon(\eta_{2},k,k_{f},p_{f}) \\
&+~g_{A}~g_{V}~(\eta_{1}.p_{f})\big(-(k.k_{i}) (k.p_{f})^2-4~(k.k_{i}) (k.p_{f}) (k.p_{i})+(k.k_{i}) (k.p_{i})^2\big)~\epsilon(\eta_{2},k,k_{f},p_{i}) \\
&+~g_{A}~g_{V}\big(-(\eta_{1}.p_{i}) (k.k_{f}) (k.p_{f})^2+4 (\eta_{1}.p_{i}) (k.k_{f}) (k.p_{f}) (k.p_{i})+2 (\eta_{1}.k_{f})(k.p_{f})^2 (k.p_{i})\\
&+(\eta_{1}.p_{i})(k.k_{f}) (k.p_{i})^2-2 (\eta_{1}.k_{f}) (k.p_{f}) (k.p_{i})^2\big)~\epsilon(\eta_{2},k,k_{i},p_{f})+g_{A}~g_{V}~\big(-(\eta_{1}.p_{f}) (k.k_{f})\\
&\times (k.p_{f})^2+2 (\eta_{1}.k_{f}) (k.p_{f})^3 -4 (\eta_{1}.p_{f}) (k.k_{f}) (k.p_{f})(k.p_{i})+(\eta_{1}.p_{f}) (k.k_{f}) (k.p_{i})^2\big)~\epsilon(\eta_{2},k,k_{i},p_{i})\\
&-~g_{A}~g_{V}(\eta_{1}.k_{f}) (k.k_{i}) ((k.p_{f})^2 + (k.p_{i})^2)~\epsilon(\eta_{2},k,p_{f},p_{i}) \big],
\end{split} 
\end{equation}
\begin{equation}
\begin{split}
D=&\dfrac{16 e}{(k.p_{f})^2(k.p_{i})^2}\Big[(k.p_{f})(k.p_{i}) \Big(2 |\textbf{a}|(\eta_{1}.p_{i}) ((g_{A} - g_{V})^2(k_{i}.p_{f})(k.k_{f})+(g_{A} + g_{V})^2(k_{f}.p_{f})(k.k_{i}))\\
&\times(k.p_{f})+2 (|\textbf{a}|(\eta_{1}.p_{f})(g_{A} + g_{V})^2(k_{i}.p_{i})(k.k_{f})+|\textbf{a}|(\eta_{1}.p_{f})(g_{A} - g_{V})^2(k_{f}.p_{i})(k.k_{i}) -|\textbf{a}|\\
&\times(\eta_{1}.k_{f})((g_{A} - g_{V})^2 (k_{i}.p_{f})+(g_{A} + g_{V})^2(k_{i}.p_{i})) (k.p_{f})) (k.p_{i}) +|\textbf{a}|^{2} e^2 (k.k_{i})(2|\textbf{a}|(\eta_{1}.p_{f})\\
&\times(g_{A}^2 + g_{V}^2)(k.k_{f})+2 |\textbf{a}|(\eta_{1}.p_{i})(g_{A}^2 + g_{V}^2)(k.k_{f})-|\textbf{a}|(\eta_{1}.k_{f})(g_{A} + g_{V})^2(k.p_{f})-|\textbf{a}|(\eta_{1}.k_{f})\\
&\times(g_{A} - g_{V})^2 (k.p_{i}))\Big)+|\textbf{a}|(k.p_{f})(k.p_{i}) \big((k.p_{f})+ (k.p_{i})\big)\Big(|\textbf{a}|^{2} e^2 (g_{A}^2 + g_{V}^2) + 2 g_{V}^2 (m^2 - (p_{i}.p_{f}))\\
& - 2 g_{A}^2 (m^2 + (p_{i}.p_{f}))\Big)\epsilon(\eta_{2},k,k_{f},k_{i})+g_{A} g_{V}|\textbf{a}|(k.p_{i}) \Big(2 (k_{i}.p_{i})(k.p_{f})((k.p_{f})-2 (k.p_{i}))+|\textbf{a}|^{2} e^2 \\
&\times(k.k_{i})((k.p_{f}) - (k.p_{i}))\Big)\epsilon(\eta_{2},k,k_{f},p_{f})+g_{A} g_{V}|\textbf{a}|\Big(|\textbf{a}|^{2}e^2 (k.k_{i})(k.p_{f})^2-|\textbf{a}|^{2}e^2 (k.k_{i})(k.p_{f})\\
&\times(k.p_{i})+2 (k_{i}.p_{f})(k.p_{f})^2(k.p_{i})\Big)\epsilon(\eta_{2},k,k_{f},p_{i}) +g_{A} g_{V}|\textbf{a}|\Big(|\textbf{a}|^{2} e^2  (k.k_{f})(k.p_{f})(k.p_{i})+2(k_{f}.p_{i})\\
&\times(k.p_{f})^2 (k.p_{i})-|\textbf{a}|^{2} e^2(k.k_{f})(k.p_{i})^2-4(k_{f}.p_{i})(k.p_{f}) (kp_{i})^2 \Big)\epsilon(\eta_{2},k,k_{i},p_{f})+g_{A} g_{V} |\textbf{a}|\Big(|\textbf{a}|^{2} e^2 (k.k_{f})\\
&\times(k.p_{f})^2 -|\textbf{a}|^{2} e^2 (k.k_{f})(k.p_{f})(k.p_{i})+2(k_{f}.p_{f})(k.p_{f})^2 (k.p_{i})\Big)\epsilon(\eta_{2},k,k_{i},p_{i})-2g_{A} g_{V}|\textbf{a}|\Big( |\textbf{a}|^{2} e^2 \\
&\times (k.k_{f})(k.k_{i})(k.p_{f})+ |\textbf{a}|^{2} e^2 (k.k_{f})(k.k_{i})(k.p_{i})-2(k_{i}.k_{f})(k.p_{f})^2 (k.p_{i})\Big)\epsilon(\eta_{2},k,p_{f},p_{i})+|\textbf{a}|\Big(2 g_{A}^2\\
&\times (k.p_{f})(k.p_{i})^3 + 2 g_{V}^2(k.p_{f})(k.p_{i})^3 \Big)\epsilon(\eta_{2},k_{f},k_{i},p_{f})+|\textbf{a}|\Big(2 g_{A}^2 (k.p_{f})^3(k.p_{i})+2 g_{V}^2 (k.p_{f})^3 (k.p_{i})\Big)\\
&\times\epsilon(\eta_{2},k_{f},k_{i},p_{i})-2 g_{A} g_{V}|\textbf{a}| (k.k_{i})(k.p_{f})^2 (k.p_{i}) \epsilon(\eta_{2},k_{f},p_{f},p_{i})-2|\textbf{a}| g_{A} g_{V} (k.k_{f})(k.p_{f})^2 (k.p_{i})\\
&\times \epsilon(\eta_{2},k_{i},p_{f},p_{i})-2g_{A}^2\Big(|\textbf{a}| (\eta_{2}.p_{i})(k.p_{f})(k.p_{i})^2+|\textbf{a}|(\eta_{2}.p_{i})(k.p_{f})(k.p_{i})^2 \Big)\epsilon(k,k_{f},k_{i},p_{f})-2g_{A}^2\\
&\times\Big(|\textbf{a}|(\eta_{2}.p_{f})(k.p_{f})^2(k.p_{i})- |\textbf{a}|(\eta_{2}.p_{f}) (k.p_{f})^2 (k.p_{i})\Big)\epsilon(k,k_{f},k_{i},p_{i})+2 |\textbf{a}|(\eta_{2}.k_{f}) g_{A} g_{V}\\
&\times (k.p_{f})^2 (k.p_{i})\epsilon(k,k_{i},p_{f},p_{i}) 
\Big],
\end{split}
\end{equation}
\begin{equation}
\begin{split}
E=&\dfrac{-16 e }{(k.p_{f})^2(k.p_{i})^2}\Big[(k.p_{f})(k.p_{i}) \Big(-2|\textbf{a}| (\eta_{1}.p_{i})((g_{A}-g_{V})^2 (k_{i}.p_{f})(k.k_{f})+(g_{A} + g_{V})^2(k_{f}.p_{f})\\&\times(k.k_{i}))(k.p_{f}) +2(-|\textbf{a}|(\eta_{1}.p_{f})(g_{A} + g_{V})^2 (k_{i}.p_{i})(k.k_{f})-|\textbf{a}|(\eta_{1}.p_{f})(g_{A} - g_{V})^2(k_{f}.p_{i}) (k.k_{i})\\
&+|\textbf{a}|(\eta_{1}.k_{f})((g_{A} - g_{V})^2(k_{i}.p_{f})+(g_{A} + g_{V})^2(k_{i}.p_{i}))(k.p_{f}))(k.p_{i})+|\textbf{a}|^{2} e^2 (k.k_{i})(-2 |\textbf{a}|(\eta_{1}.p_{f})\\
&\times(g_{A}^2 + g_{V}^2)(k.k_{f})-2|\textbf{a}| (\eta_{1}.p_{i}) (g_{A}^2+g_{V}^2)(k.k_{f})+|\textbf{a}|(\eta_{1}.k_{f})(g_{A} + g_{V})^2(k.p_{f}) +|\textbf{a}|(\eta_{1}.k_{f})\\
&\times(g_{A} - g_{V})^2 (k.p_{i}))\Big)+(k.p_{f})(k.p_{i})((k.p_{f})+|\textbf{a}|(k.p_{i}))\Big(|\textbf{a}|^{2} e^2(g_{A}^2+g_{V}^2)+ 2 g_{V}^2 (m^2 - (p_{i}.p_{f}))\\
&-2 g_{A}^2 (m^2 + (p_{i}.p_{f}))\Big) \epsilon(\eta_{2},k,k_{f},k_{i}) +g_{A} g_{V}|\textbf{a}| \Big(2 (k_{i}.p_{i})(k.p_{f})((k.p_{f}) - 2 (k.p_{i}))+|\textbf{a}|^{2} e^2 (k.k_{i})\\
&\times((k.p_{f})-(k.p_{i}))\Big)(k.p_{i})\epsilon(\eta_{2},k,k_{f},p_{f})+ g_{A} g_{V}|\textbf{a}|\Big(|\textbf{a}|^{2} e^2 (k.k_{i}) (k.p_{f})^2-|\textbf{a}|^{2} e^2 (k.k_{i})(k.p_{f})\\
&\times(k.p_{i})+2(k_{i}.p_{f})(k.p_{f})^2(k.p_{i})\Big)\epsilon(\eta_{2},k,k_{f},p_{i}) +g_{A} g_{V}|\textbf{a}|\Big(|\textbf{a}|^{2} e^2(k.k_{f}) (k.p_{f})(k.p_{i})+2 (k_{f}.p_{i})\\\
&\times(k.p_{f})^2 (k.p_{i})-|\textbf{a}|^{2} e^2 (k.k_{f}) (k.p_{i})^2-4 (k_{f}.p_{i})(k.p_{f})(k.p_{i})^2\Big)\epsilon(\eta_{2},k,k_{i},p_{f})+g_{A} g_{V}|\textbf{a}|\Big(|\textbf{a}|^{2} e^2 \\
&\times(k.k_{f})(k.p_{f})^2-|\textbf{a}|^{2} e^2 (k.k_{f})(k.p_{f})(k.p_{i})+2 (k_{f}.p_{f})(k.p_{f})^2(k.p_{i}) \Big)\epsilon(\eta_{2},k,k_{i},p_{i})+2g_{A} g_{V}|\textbf{a}|\\
&\times\Big( |\textbf{a}|^{2} e^2 (k.k_{f})(k.k_{i})(k.p_{f})-|\textbf{a}|^{2} e^2 (k.k_{f})(k.k_{i})(k.p_{i})+2(k_{i}.k_{f})(k.p_{f})^2(k.p_{i}) \Big)\epsilon(\eta_{2},k,p_{f},p_{i})\\
&+2(k.p_{f})(k.p_{i})^3|\textbf{a}|\big( g_{A}^2 + g_{V}^2\big)\epsilon(\eta_{2},k_{f},k_{i},p_{f})+2(k.p_{f})^3(k.p_{i})|\textbf{a}|\big(g_{A}^2+g_{V}^2\big)\epsilon(\eta_{2},k_{f},k_{i},p_{i})\\
&-2 g_{A} g_{V}|\textbf{a}|(k.k_{i})(k.p_{f})^2(k.p_{i})\epsilon(\eta_{2},k_{f},p_{f},p_{i})-2 g_{A} g_{V}|\textbf{a}| (k.k_{f})(k.p_{f})^2(k.p_{i})\epsilon(\eta_{2},k_{i},p_{f},p_{i})\\
&-2|\textbf{a}|(\eta_{2}.p_{i})(k.p_{f})(k.p_{i})^2 \big(g_{A}^2+ g_{V}^2 \Big)\epsilon(k,k_{f},k_{i},p_{f})-2 |\textbf{a}|(\eta_{2}.p_{f})(k.p_{f})^2 (k.p_{i})\big(g_{A}^2 + g_{V}^2 \big)\\
&\times\epsilon(k,k_{f},k_{i},p_{i})+2 |\textbf{a}|(\eta_{2}.k_{f}) g_{A} g_{V} (k.p_{f})^2(k.p_{i})\epsilon(k,k_{i},p_{f},p_{i}) \Big],
\end{split}
\end{equation}
\begin{equation}
\begin{split}
F=&\dfrac{32 e^2 (k.k_{i}) }{(k.p_{f})(k.p_{i})}\Big[2|\textbf{a}|^2(\eta_{1}.p_{f})(\eta_{1}.p_{i})(g_{A}^2 + g_{V}^2)(k.k_{f})-2|\textbf{a}|^2(\eta_{2}.p_{f})(\eta_{2}.p_{i})(g_{A}^2 + g_{V}^2)(k.k_{f})\\
&-|\textbf{a}|^2((\eta_{1}.k_{f})(\eta_{1}.p_{i})-(\eta_{2}.k_{f})(\eta_{2}.p_{i}))(g_{A} +g_{V})^2(k.p_{f})-|\textbf{a}|^2(\eta_{1}.k_{f})(\eta_{1}.p_{f})(g_{A} - g_{V})^2(k.p_{i})\\
&+|\textbf{a}|^2(\eta_{2}.k_{f})(\eta_{2}.p_{f})(g_{A} - g_{V})^2(k.p_{i})\Big],
\end{split}
\end{equation}
where, for all 4-vectors $a$, $b$, $c$ and $d$, we have
\begin{equation}
\begin{split}
\epsilon(a,b,c,d)= \epsilon_{\mu\nu\rho\sigma} a^{\mu}b^{\nu}c^{\rho}d^{\sigma}.
\end{split}
\end{equation}
We notice that, in the two coefficients $A$ and $F$, there is no appearance of the antisymmetric tensors. This obviously means that they have been completely contracted. However, the other coefficients $B$, $C$, $D$ and $E$ contained different noncontracted tensors. To calculate analytically these tensors, we recall here that we used the Grozin convention
\begin{equation}
\begin{split}
\epsilon_{0123}=1, 
\end{split}
\end{equation}
which means that $\epsilon_{\mu\nu\rho\sigma}=1~(-1)$ for an even (odd) permutation of the Lorentz indices and $\epsilon_{\mu\mu\rho\sigma}=0$ otherwise. For instance, we give below how we have calculated one of these tensors, for example in the coefficient $D$
\begin{equation}
\begin{split}
\epsilon(\eta_{2},k,k_{f},k_{i})=& \epsilon_{\mu\nu\rho\sigma} \eta_{2}^{\mu}k^{\nu}k_{f}^{\rho}k_{i}^{\sigma},\\
=& \omega\big(\epsilon_{2013}k_{f}^{1}k_{i}^{3}+\epsilon_{2031}k_{f}^{3}k_{i}^{1}\big)+\omega\big(\epsilon_{2301}k_{f}^{0}k_{i}^{1}+\epsilon_{2310}k_{f}^{1}k_{i}^{0}\big),\\
=&-2\omega E_{i}\big(|\textbf{q}_{i}|\cos(\phi_{i}) \sin(\theta_{i})-|\textbf{q}_{f}|\cos(\phi_{f}) \sin(\theta_{f})\big).
\end{split}
\end{equation}
One can verify that in the case where $|\textbf{a}|\rightarrow 0$ and $n=0$, which implies $z=0$, all the terms that contribute to Eq.~(\ref{mfiappendix}) vanish except for the term multiplied by the coefficient $A$, since $J_{n}(0)=\delta_{0n}$ where $\delta_{0n}$ is the Kronecker's delta satisfying $\delta_{0n}=1$ if $n=0$ and $\delta_{0n}=0$ otherwise. Looking at the expression of the coefficient $A$ in Eq.~(\ref{coeffA}), we find that in this case it simplifies to match the expression obtained in the absence of the laser field given in Eq.~(\ref{restracefree}).


\begin{thebibliography}{99}
\bibitem{rutherford} E. Rutherford, The scattering of $\alpha$ and $\beta$ particles by matter and the structure of the atom, Phil. Mag. \textbf{21}, 669 (1911).
\bibitem{higgs} S. Chatrchyan \textit{et al.} (CMS collaboration), Observation of a new boson at a mass of 125 GeV with the CMS experiment at the LHC, Phys. Lett. B \textbf{716}, 30 (2012).
\bibitem{laser} H. Kiriyama \textit{et al.}, High-contrast high-intensity repetitive petawatt laser, Opt. Lett. \textbf{43}, 2595 (2018).
\bibitem{laser1} C. N. Danson \textit{et al.}, Petawatt and exawatt class lasers worldwide, High Power Laser Sci. Eng. \textbf{7}, e54 (2019).
\bibitem{laser2} J. W. Yoon, C. Jeon, J. Shin, S. K. Lee, H. W. Lee, I. W. Choi, H. T. Kim, J. H. Sung, and C. H. Nam, Achieving the laser intensity of $5.5\times10^{22}~\text{W/cm}^{2}$ with a wavefront-corrected multi-PW laser, Opt. Express \textbf{27}, 20412 (2019).
\bibitem{nobel2018} D. Strickland, Nobel Lecture: Generating high-intensity ultrashort optical pulses, Rev. Mod. Phys. \textbf{91},  030502 (2019).
\bibitem{mourou} D. Strickland and G. Mourou, Compression of amplified chirped optical pulses, Opt. Commun. \textbf{56}, 219 (1985).
\bibitem{qed0} F. C. Vélez, J. Z. Kaminski, and K. Krajewska, Electron scattering processes in non-monochromatic and relativistically intense laser fields, Atoms \textbf{7}, 34 (2019).
\bibitem{hartin} A. Hartin, Strong field QED in lepton colliders and electron/laser interactions, Int. J. Mod. Phys. A \textbf{33}, 1830011 (2018).
\bibitem{qed1} A. Di Piazza, C. M\"{u}ller, K. Z. Hatsagortsyan, and C. H. Keitel, Extremely high-intensity laser interactions with fundamental quantum systems, Rev. Mod. Phys. \textbf{84}, 1177 (2012).
\bibitem{qed2}  F. Ehlotzky, K. Krajewska, and J. Z. Kamiński, Fundamental processes of quantum electrodynamics in laser fields of relativistic power, Rep. Prog. Phys. \textbf{72}, 046401 (2009).
\bibitem{mouslih} S. Mouslih, M. Jakha, S. Taj, B. Manaut, and E. Siher, Laser-assisted pion decay, Phys. Rev. D \textbf{102}, 073006 (2020).
\bibitem{jakha} M. Jakha, S. Mouslih, S. Taj, and B. Manaut, Laser effect on the final products of $Z$-boson decay, Laser Phys. Lett. \textbf{18}, 016002 (2021) [arXiv:2010.14401]. 
\bibitem{muon1} A.-H. Liu, S.-M. Li, and J. Berakdar, Laser-assisted muon decay, Phys. Rev. Lett. \textbf{98}, 251803 (2007).
\bibitem{muon2} D. A. Dicus, A. Farzinnia, W. W. Repko, and T. M. Tinsley, Muon decay in a laser field, Phys. Rev. D \textbf{79}, 013004 (2009).
\bibitem{muon3} A. Farzinnia, D. A. Dicus, W. W. Repko, and T. M. Tinsley, Muon decay in a linearly polarized laser field, Phys. Rev. D \textbf{80}, 073004 (2009).
\bibitem{atom} B. Manaut, S. Taj, and M. El Idrissi, Relativistic elastic scattering of hydrogen atom by positron impact in a circularly polarized laser field, Can. J. Phys. \textbf{91}, 696 (2013).
\bibitem{atom1} A. Makhoute, D. Khalil, and I. Ajana, Laser-Assisted (e, 2e) Collisions in the Symmetric/Asymmetric Coplanar Geometry, Atoms \textbf{7}, 40 (2019). 
\bibitem{atom2} Y. Attaourti, B. Manaut, and A. Makhoute, Relativistic electronic dressing in laser-assisted electron-hydrogen elastic collisions, Phys. Rev. A \textbf{69}, 063407 (2004).
\bibitem{atom3} Y. Attaourti and S. Taj, Relativistic electronic dressing in laser-assisted ionization of atomic hydrogen by electron impact, Phys. Rev. A \textbf{69}, 063411 (2004).
\bibitem{bulanov} S. V. Bulanov, T. Zh. Esirkepov, D. Habs, F. Pegoraro, and T. Tajima, Relativistic laser-matter interaction and relativistic laboratory astrophysics, Eur. Phys. J. D \textbf{55}, 483 (2009).
\bibitem{salamin} Y. I. Salamin, S. X. Hu, K. Z. Hatsagortsyan, and C. H. Keitel, Relativistic high-power laser-matter interactions, Phys. Rep. \textbf{427}, 41 (2006).
\bibitem{test1} T. Kumita \textit{et al.}, Observation of the nonlinear effect in relativistic Thomson scattering of electron and laser beams, Laser Phys. \textbf{16}, 267 (2006). 
\bibitem{test2} D. L. Burke \textit{et al.}, Positron production in multiphoton light-by-light scattering, Phys. Rev. Lett. \textbf{79}, 1626 (1997).
\bibitem{test3} C. Bula \textit{et al.}, Observation of nonlinear effects in Compton scattering,  Phys. Rev. Lett. \textbf{76}, 3116 (1996).
\bibitem{bai2012} L. Bai, M.-Y. Zheng, and B.-H. Wang, Multiphoton processes in laser-assisted scattering of a muon neutrino by an electron, Phys. Rev. A \textbf{85}, 013402 (2012).
\bibitem{vilain} P. Vilain \textit{et al.} (CHARM II Collaboration), Precision measurement of electroweak parameters from the scattering of muon-neutrinos on electrons, Phys. Lett. B \textbf{335}, 246 (1994).
\bibitem{hasert} F. J. Hasert \textit{et al.} (Gargamelle Collaboration), Search for elastic muon-neutrino electron scattering, Phys. Lett. B \textbf{46}, 121 (1973).
\bibitem{tomalak} O. Tomalak and R. J. Hill, Theory of elastic neutrino-electron scattering, Phys. Rev. D \textbf{101}, 033006 (2020).
\bibitem{marciano} W. J. Marciano and Z. Parsa, Neutrino-electron scattering theory, J. Phys. G: Nucl. Part. Phys. \textbf{29}, 2629 (2003).
\bibitem{tinsley} T. M. Tinsley, Pair production with neutrinos and high-intensity laser fields, Phys. Rev. D \textbf{71}, 073010 (2005).
\bibitem{dicus} D. A. Dicus, W. W. Repko, and T. M. Tinsley, Pair production with neutrinos in an intense background magnetic field, Phys. Rev. D \textbf{76}, 025005 (2007).
\bibitem{pal} K. Bhattacharya and P. B. Pal, Neutrinos and magnetic fields: a short review, Proc. Indian Natl. Sci. Acad., Part A \textbf{70}, 145 (2004). 
\bibitem{pdg2020} P. A. Zyla \textit{et al.} (Particle Data Group), Review of Particle Physics, Prog. Theor. Exp. Phys. \textbf{2020}, 083C01 (2020).
\bibitem{greiner} W. Greiner and B. M\"{u}ller, \textit{Gauge Theory of Weak Interactions}, 4th ed. (Springer, Berlin, 2009).
\bibitem{landau} V. B. Berestetskii, E. M. Lifshitz, and L. P. Pitaevskii, \emph{Quantum Electrodynamics}, 2nd ed. (Butterworth-Heinemann, Oxford, 1982).
\bibitem{volkov} D. M. Volkov, On a class of solutions of the Dirac equation, Z. Phys. \textbf{94}, 250 (1935).
\bibitem{feyncalc1} R. Mertig, M. B\"{o}hm, and A. Denner, Feyn Calc - Computer-algebraic calculation of Feynman amplitudes, Comput. Phys. Commun. \textbf{64}, 345 (1991).
\bibitem{feyncalc2} V. Shtabovenko, R. Mertig, and F. Orellana,  New developments in FeynCalc 9.0,  Comput. Phys. Commun. \textbf{207}, 432 (2016). 
\bibitem{feyncalc3} V. Shtabovenko, R. Mertig, and F. Orellana, FeynCalc 9.3: New features and improvements, Comput. Phys. Commun. \textbf{256}, 107478 (2020).
\bibitem{kajita2010} T. Kajita, Atmospheric neutrinos and discovery of neutrino oscillations, Proc. Jpn. Acad., Ser. B \textbf{86}, 303 (2010).
\bibitem{kajita2006} T. Kajita, Discovery of neutrino oscillations, Rep. Prog. Phys. \textbf{69}, 1607 (2006).
\bibitem{fukuda98} Y. Fukuda \textit{et al.} (Super-Kamiokande Collaboration), Evidence for oscillation of atmospheric neutrinos, Phys. Rev. Lett. \textbf{81}, 1562 (1998).
\bibitem{danby62} G. Danby, J. M. Gaillard, K. A. Goulianos, L. M. Lederman, N. B. Mistry, M. Schwartz, and J. Steinberger, Observation of high-energy neutrino reactions and the existence of two kinds of neutrinos, Phys. Rev. Lett. \textbf{9}, 36 (1962).
\bibitem{nobel1988} The Nobel Prize in Physics 1988. \url{https://www.nobelprize.org/prizes/physics/1988/summary/}.
\bibitem{aliu2005} E. Aliu \textit{et al.} (The K2K Collaboration), Evidence for muon neutrino oscillation in an accelerator-based experiment, Phys. Rev. Lett. \textbf{94}, 081802 (2005).
\bibitem{assamagan} K. Assamagan \textit{et al.}, Upper limit of the muon-neutrino mass and charged-pion mass from momentum analysis of a surface muon beam, Phys. Rev. D \textbf{53}, 6065 (1996).
\bibitem{katrin} G.-y. Huang, W. Rodejohann, and S. Zhou, Effective neutrino masses in KATRIN and future tritium beta-decay experiments, Phys. Rev. D \textbf{101}, 016003 (2020).
\bibitem{formaggio2013} J. A. Formaggio and G. P. Zeller, From eV to EeV: neutrino cross-sections across energy scales, Rev. Mod. Phys. \textbf{84}, 1307 (2012).
\bibitem{baker1989} N. J. Baker \textit{et al.}, Measurement of muon-neutrino-electron elastic scattering in the Fermilab 15-foot bubble chamber, Phys. Rev. D \textbf{40}, 2753 (1989).
\bibitem{faissner1978} H. Faissner \textit{et al.}, Measurement of muon-neutrino and -antineutrino scattering off electrons, Phys. Rev. Lett. \textbf{41}, 213 (1978).
\bibitem{lasercp} C. L. Zhong, B. Qiao, X. R. Xu, Y. X. Zhang, X. B. Li, Y. Zhang, C. T. Zhou, S. P. Zhu, and X. T. He, Intense circularly polarized attosecond pulse generation from solid targets irradiated with a two-color linearly polarized laser, Phys. Rev. A \textbf{101}, 053814 (2020).
\bibitem{szymanowski} C. Szymanowski, V. Véniard, R. Ta\"{i}eb, A. Maquet and C. H. Keitel, Mott scattering in strong laser fields, Phys. Rev. A \textbf{56}, 3846 (1997).
\bibitem{sumrule} N. M. Kroll and K. M. Watson, Charged-particle scattering in the presence of a strong electromagnetic wave, Phys. Rev. A \textbf{8}, 804 (1973).
\bibitem{mouslih1} S. Mouslih, M. Jakha, I. Dahiri, S. Taj, B. Manaut, and E. Siher, New phenomena in laser-assisted leptonic decays of the negatively charged boson $W^{-}$, arXiv:2101.00224.
\bibitem{imane} I. Dahiri, M. Jakha, S. Mouslih, B. Manaut, and S. Taj, Elastic electron-proton scattering in the presence of a circularly polarized laser field, arXiv:2102.00722.
\bibitem{hrour} E. Hrour, S. Taj, A. Chahboune, and B. Manaut, Relativistic proton-impact excitation of hydrogen atom in the presence of intense laser field, Can. J. Phys. \textbf{94}, 645 (2016).
\bibitem{taj} S. Taj, B. Manaut, M. El Idrissi, and L. Oufni, Laser-assisted semi-relativistic excitation of atomic hydrogen by electronic impact, Chin. J. Phys. \textbf{49}, 1164 (2011).
\bibitem{geltman} S. Geltman, Low-energy laser-assisted electron-helium collisions, Phys. Rev. A \textbf{55}, 3755 (1997). 
\bibitem{geltman1} S. Geltman, Laser-assisted collisions: The Kroll-Watson formula and bremsstrahlung theory, Phys. Rev. A \textbf{53}, 3473 (1996).
\bibitem{kaminski} J. Z. Kaminski, Relativistic generalisation of the Kroll-Watson formula, J. Phys. A: Math. Gen. \textbf{18}, 3365 (1985).
\bibitem{holmes} B. Wallbank and J.K. Holmes, Laser-assisted elastic electron
scattering from helium, Can. J. Phys. \textbf{79}, 1237 (2001).
\bibitem{kurilin2004} A. V. Kurilin, Leptonic decays of the $W$ boson in a strong electromagnetic field, Phys. Atom. Nucl. \textbf{67}, 2095 (2004).
\bibitem{kurilin88} V. Ch. Zhukovsky and A. V. Kurilin, $W$-boson creation in an intense electromagnetic field and the $W$ contribution to the radiative shift of the electron mass, Yad. Fiz. \textbf{48}, 179 (1988) [Sov. J. Nucl. Phys. \textbf{48}, 114 (1988)].
\end{thebibliography}
\end{document}